 \documentclass[final,3p,times]{elsarticle}

\usepackage{graphicx}
\usepackage{amssymb}
\usepackage{tabularx}
\usepackage{newtxmath}
\usepackage{mathbbol}

\usepackage{lineno}

\usepackage{hyperref}

\newcommand{\beq}{\begin{equation}}
\newcommand{\eeq}{\end{equation}}

\def\scas  #1{\mbox{{\scriptsize{${\rm{#1}}$}}}{}}

\usepackage{comment}
\usepackage{amsmath}
\usepackage{lipsum}
\usepackage{graphics}
\usepackage[dvipsnames]{xcolor}
\usepackage{ulem}

\journal{Arxiv}

\begin{document}

\begin{frontmatter}


\title{\textcolor{black}{Predicting the Mechanical Properties of Biopolymer Gels Using Neural Networks Trained on Discrete Fiber Network Data}}



\author{Yue Leng$^{1}$, \textcolor{black}{Vahidullah Tac$^2$}, Sarah Calve$^{1,2}$ and Adrian B Tepole$^{1,3}$ }

\address{$^1$Weldon School of Biomedical Engineering, Purdue University, West Lafayette, IN, USA\\ $^2$Department of Mechanical Engineering, Colorado University, Boulder, CO, USA\\ $^3$School of Mechanical Engineering, Purdue University, West Lafayette, IN, USA}

\begin{abstract}
\textcolor{black}{Biopolymer gels, such as those made out of fibrin or collagen, are widely used in tissue engineering applications and biomedical research}. Moreover, fibrin naturally assembles into gels \textit{in vivo} during wound healing and thrombus formation. The macroscaole properties of fibrin and other \textcolor{black}{biopolymer} gels are dictated by the response of a microscale fiber network. Hence, accurate description of \textcolor{black}{biopolymer} gels can be achieved using representative volume elements (RVE) that explicitly model the discrete fiber networks of the microscale. These RVE models, however, cannot be efficiently used to model the macroscale due to the challenges and computational demands of multiscale coupling. Here, we propose the use of an artificial, fully connected neural network (FCNN) to efficiently capture the behavior of the RVE models. The FCNN was trained on 1100 fiber networks subjected to 121 biaxial deformations. The stress data from the RVE, together with the total energy on the fibers and the condition of incompressibility of the surrounding matrix, were used to determine the derivatives of an unknown strain energy function with respect to the deformation invariants. During training, the loss function was modified to ensure convexity of the strain energy function and symmetry of its Hessian. A general FCNN model was coded into a user material subroutine (UMAT) in the software Abaqus. The UMAT implementation takes in the structure and parameters of an arbitrary FCNN as material parameters from the input file. The inputs to the FCNN include the first two isochoric invariants of the deformation. The FCNN outputs the derivatives of the strain energy with respect to the isochoric invariants. In this work, the FCNN trained on the discrete fiber network data was used in finite element simulations of \textcolor{black}{biopolymer} gels using our UMAT. We anticipate that this work will enable further integration of machine learning tools with computational mechanics. It will also improve computational modeling of biological materials characterized by a multiscale structure.
\end{abstract}


\end{frontmatter}

\section*{Introduction}\label{motiv}

It is well known that the biomechanical macroscopic behavior of both native and engineered tissues is largely determined by the properties of the underlying microstructural components \cite{sander2009image}. Therefore, to allow for a better design of engineered tissues and to increase our fundamental understanding of the mechanical behavior of soft tissue, it is necessary to investigate this multiscale coupling. \textcolor{black}{In this paper we develop a model to describe biopolymer gels, and use fibrin as an example.} Fibrin is an important extracellular matrix (ECM) component in the body, providing structural integrity to various tissues \cite{lai2012microstructural,mol2005fibrin}. It plays a critical role in wound healing and thrombus fate \cite{schlag1986importance,laurens2006fibrin}. In addition, it is also a common scaffold material used in tissue engineering \cite{li2015fibrin}. \textcolor{black}{Collagen gels are also widely used in tissue engineering applications and self-assemble to form fibrillar structures \textit{in vitro} \cite{hadi2012simulated,sander2009image}.} In general, the fiber network structure of \textcolor{black}{biopolymer} gels can be described by variables such as the fiber diameter and length, the number of branch points, and the volume fraction \cite{stylianopoulos2007volume}. 

Due to the inherent multiscale nature of \textcolor{black}{biopolymer} gels, with the macroscopic dimension described at the centimeter length scale and the underlying fiber network defined at the micrometer length scale, our understanding of the macroscale mechanics needs to be informed by the microscale mechanics of the fiber network. To enable the multiscale description, models of discrete fiber networks (DFN) in a representative volume element (RVE) have often been proposed \cite{li20163d,hadi2012simulated}. These micromechanical models, based on volume averaging theories, assume that the stress and strains at the macroscopic scale are volume averages of the corresponding microscale fields in the RVE, and thus provide the linkage between the fiber-level mechanics and the tissue-level behaviors \cite{stylianopoulos2007volume,agoram2001coupled,driessen2005structural}. DFN models are able to predict the mechanical behavior of three-dimensional \textcolor{black}{biopolymer} networks, \textcolor{black}{e.g. fibrin and collagen gels}. The volume-averaging homogenization that satisfies the Hill-Mandel condition can then be used as a constitutive model at the macroscale, where the standard balance of linear momentum within continuum mechanics can be solved to determine the mechanical response of the gel \cite{geers2010multi,fish2008mathematical}. Moreover, at the macroscale, hyperelasticity is a common assumption that is used to describe soft tissue mechanics  \cite{natali2006hyperelastic,freed2010hypoelastic}. Numerically, the macroscale response can be efficiently modeled within a nonlinear finite element framework. The standard coupling between the macroscale and microscale descriptions, however, entails the simulation of individual RVEs associated with each of the integration points of the finite element simulation. This strategy of nesting DFN models within the finite element framework for large scale heterogeneous structures is computationally prohibitive. Not surprisingly, their use has been limited \cite{alber2019integrating,reimann2019modeling}.

A myriad of strategies for model order reduction have been implemented to predict the mechanical properties of multiscale materials that balance computational cost and accuracy \cite{bhattacharjee2016nonlinear,kerfriden2013partitioned,liu2015statistical,michel2016model,oliver2017reduced}. In particular, machine learning (ML) methods have been extended to various problems of mechanics modeling and multiscale simulations, e.g., modeling macroscopic material behavior by stress-strain curves obtained from micromechanical simulations \textcolor{black}{\cite{reimann2019modeling,yang2020prediction,vlassis2020geometric}}; approximating macroscopic stress and elasticity tensor components directly from data \cite{le2015computational,lejeune2020exploring,lejeune2020mechanical}; capturing the multiscale hydro-mechanical coupling effect of porous media with pores of varying sizes \cite{wang2018multiscale}, and predicting the stress-strain behavior of soft tissues including growth and remodeling \cite{costabal2019multiscale,peirlinck2019using,lee2020propagation}.

The neural network method is one example of a ML technique that is capable of extracting relationships from data given a sufficient amount of training data. This method is efficient in automatically discovering and capturing the underlying complex high-dimensional mapping from the feature vector input to the desired output without the need of manually deriving specific functional forms \cite{peng2020multiscale,tepole2020special}. 

In this paper, we propose a novel multiscale modeling approach that employs fully connected neural networks (FCNN) to describe the homogenized response of DFNs. The FCNN is able to  predict the mechanical response of various microstructures under different loads. In this way, the trained FCNN represents a new macroscopic constitutive relation. To develop the FCNN metamodel, we first generated training data by creating a large number of DFN microstructures and subjecting the RVEs to various biaxial loading conditions. Derivatives of the strain energy with respect to the strain invariants were obtained by creating a Gaussian process (GP) surrogate for the strain energy further constrained on the observed stress data. Using the values of the derivatives of the strain energy, we trained a FCNN model to predict the mechanical properties (e.g., stress and elasticity tensor) of a gel based on its microstructure. Notably, the convexity of the strain energy, which is an essential constraint to achieve physically realistic  and numerically stable simulations, was considered in the training process by specifically defining the loss function to enforce these constraints. Following training, we tested the FCNN by comparing the prediction against the ground truth DFN model on a validation set. Finally, we implemented the FCNN in a general User Material Subroutine (UMAT) in the popular finite element package Abaqus.  

Our work demonstrates that neural networks can be trained by micromechanical simulations, which capture ECM network behavior accurately. \textcolor{black}{We use fibrin gels as a representative material that can be described with the proposed methodology}. \textcolor{black}{The architecture of the FCNN is designed to satisfy the objectivity and convexity of the material model. Objectivity is ensured because the FCNN is completely based on invariants inputs, and it outputs the strain energy and the strain energy derivatives, which are scalar (invariant) functions of the deformation. Convexity is imposed through the loss function.} \textcolor{black}{The proposed FCNN material model leads to an efficient and robust implementation into a nonlinear static equilibrium finite element solver. The finite element implementation is enabled through the invariant formulation, the convexity constraint, and the direct prediction of the derivatives of the strain energy}. We anticipate that this work will enable the widespread use of ML-driven multiscale simulations by reducing the computational cost without compromising accuracy.

\begin{figure}[h!]
\centering
\includegraphics[width=0.7\linewidth]{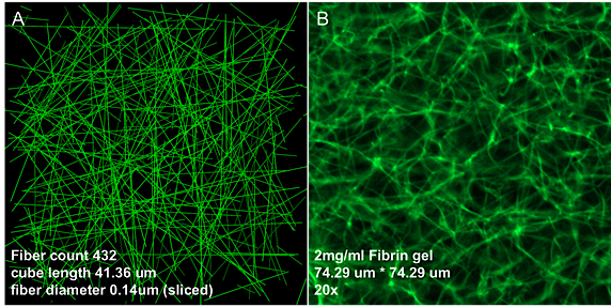}
\caption{A microscale network model (A), was used to simulate the fibrin networks (B). }
\label{fig01} 
\end{figure}

\section*{Materials and Methods}

\subsection*{Discrete fiber network models} 

In the DFN model, multiple fibers were linked to each other through cross-links, and they were allowed to stretch, compress, or rotate at these points. Bending forces were not considered during fiber deformation. The procedure for the generation of the networks was in a similar manner to previous efforts \cite{stylianopoulos2007volume,clague1997numerical}. Namely, we seeded the three-dimensional RVE with random uniformly distributed nucleation points. Each point gave rise to two segments, which grew oppositely along a randomly chosen direction sampled from a uniform distribution on the unit sphere. At each step of fiber growth, every segment was checked to decide if it had exceeded the boundaries of the cube or had collided with another segment. At the point of the identified intersection or collision, a node was introduced, and the segment stopped growing. When the growth of all the segments stopped, the algorithm of fibrillar network generation terminated. A sample image of a DFN is depicted in Figure \ref{fig01}A, alongside a confocal image of a 2mg/ml fibrin gel in Figure \ref{fig01}B.

Applying forces or deformations to the boundaries of the RVE leads to the deformation of the fibers inside of the network. Individual fibers were considered as hyperelastic and the state of mechanical equilibrium was determined by finding the minimum of the total energy in the fibers. The energy in a fiber $i$ in the network is

\begin{equation}
\label{eq01}
\Psi_f^{(i)} = \frac{k_f}{2}\left( (\lambda_f^{(i)})^2-1\right)\, , 
\end{equation}

where the stretch of the fiber $i$ is given by

\beq 
\lambda_f^{(i)} = \frac{||\mathbf{x}_1^{(i)}-\mathbf{x}_0^{(i)}||}{||\mathbf{X}_1^{(i)}-\mathbf{X}_0^{(i)}||} \, ,
\eeq 

$\mathbf{X}_0^{(i)},\mathbf{X}_1^{(i)}$ denote the reference coordinates of the nodes making up the fiber, and $\mathbf{x}_0^{(i)},\mathbf{x}_1^{(i)}$ the location of the nodes after deformation. The material parameter $k_f$ is the stiffness of the fiber. The mechanical equilibrium is obtained by searching for the deformed nodal coordinates of all the fibers such that the total energy $\sum_i \Psi^{(i)}_f$ is minimized. The boundary conditions are imposed displacements on all nodes at the boundary of the RVE. The homogenized stress on the RVE is calculated based on the averaging  \cite{stylianopoulos2007volume},

\beq
\mathbf{\sigma}^M = \langle \mathbf{\sigma}^\mu \rangle = \frac{1}{V}\int_{RVE}  \mathbf{\sigma}^\mu \, .
\eeq

 $\mathbf{\sigma}^M$ denotes the macroscale stress, which is the average of the microscale stress field $\mathbf{\sigma}^\mu$ over the RVE volume $V$. For the DFN, the stress can be written in indicial notation as

\beq 
\sigma_{ij}^M  = \frac{1}{V} \sum x_i^b R_j^b
\label{eq_sigma_micro}
\eeq 

where the sum is over the deformed coordinates of the boundary nodes, $\mathbf{x}^b$, and the corresponding reaction forces $\mathbf{R}^b$. \textcolor{black}{As noted before, the discrete fiber network models closely follow previous work in the field and the reader is referred to   \cite{stylianopoulos2007volume,sander2009image} for additional details. The code to generate the DFNs and run the equilibrium simulations used in this paper is publicly available through the Github repository listed at the end of the article}.

To generate enough training data for the FCNN, a total of \textcolor{black}{$1100$} networks were generated. To explore the different network geometries, two quantities of interest were identified as the best descriptors for microstructure: the volume fraction $\theta$, and the fiber diameter $\varphi$. The range of the model parameters was obtained to span similar networks to those reported in the literature \cite{stylianopoulos2007volume,lai2012mechanical}. Namely, the diameter $\varphi$ was in the range of $[20-500]$ nm, and the volume fraction was in the range $\theta\in[0.05, 1.0]$\%.

Following the generation of the \textcolor{black}{$1100$} different RVE microstructures, data of stress and energy for each of these RVEs under various deformation regimes were generated. Analogous to mechanical tests performed to characterize soft material mechanics, the RVEs were subjected to biaxial deformation under plane stress. Eleven stretches in the $x$ axis were selected \textcolor{black}{$\lambda_x=[1.00,1.025,1.05,1.075,1.10,1.125,1.15,1.175,1.2,1.225,1.25]$}. Similarly for the $y$ axis, we used \textcolor{black}{$\lambda_y=[1.00,1.025,1.05,1.075,1.10,1.125,1.15,1.175,1.2,1.225,1.25]$}. The deformations applied to the RVE were all the possible combinations of $\lambda_x$ and $\lambda_y$ from the values above, i.e. a total of $121$ simulations per RVE were run for the training of the ML metamodel. The transverse stretch was calculated to maintain incompressibility. While not used for training of the ML metamodel, uniaxial simulations were also explored to compare against previous reports in the literature.      

\begin{figure}
\centering
\includegraphics[width=0.99\linewidth]{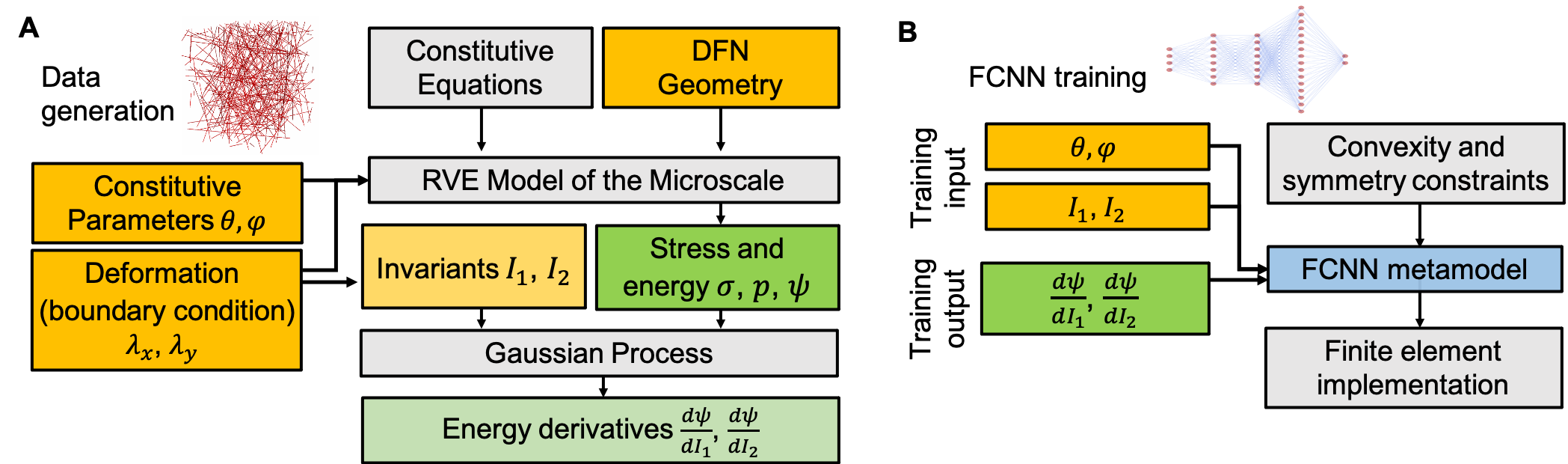}
\caption{Framework for training a ML-based microscale model. Input data was generated by sampling from microstructure $\theta, \varphi$, and deformation parameters $\lambda_x,\lambda_y$, which were post-processed to obtain the invariants $I_1,I_2$. Multiple random geometries were sampled for a given microstructure and deformation. The RVE model was evaluated to generate output stress and strain energy, which were post-processed through Gaussian process regression and optimization to obtain energy derivative outputs (A). A fully connected neural network (FCNN) was trained on a training subset of the data, and tested on a different subset of data. FCCN training was further constrained to satisfy convexity and symmetry, requirements for a stable finite element implementation (B).   }
\label{fig02} 
\end{figure}

\subsection*{Hyperelastic material behavior at the macroscale} 

In the multiscale formulation, the average stress of the homogenized RVE, $\mathbf{\sigma}^M = \langle \mathbf{\sigma}^\mu \rangle$, is a function of the average deformation gradient $\mathbf{F}^M = \langle \mathbf{F}^\mu \rangle$. In practice, the average deformation gradient known from the macroscale, $\mathbf{F}^M$, is imposed as a boundary condition to the RVE. Mechanical equilibrium of the RVE then yields the microscale field $\mathbf{\sigma}^\mu$, which is averaged to compute the corresponding stress $\mathbf{\sigma}^M$. In this way, the relationship $\mathbf{\sigma}^M(\mathbf{F}^M)$ is implied although not available explicitly, it needs instead the evaluation of the RVE model. Under the additional assumption that the macroscopic material behavior is that of a nearly incompressible hyperelastic material \cite{mihai2015finite,mihai2017microstructure}, we propose the existence of a strain energy function $\Psi$  

\beq
\Psi = \Psi_{\mathrm{iso}}(\bar{I}_1,\bar{I}_2)+\Psi_{\mathrm{vol}}(J)
\label{eq_Psi_M}
\eeq 

with the standard definition of the isochoric strain invariants and the volume change
\beq 
\bar{I}_1=\mathrm{tr}(\bar{\mathbf{b}})=\bar{\mathbf{b}}:\mathbf{I}, \; \bar{I}_2=\frac{1}{2}\left( (\mathrm{tr}(\bar{\mathbf{b}}))^2 - \mathrm{tr}(\bar{\mathbf{b}}^2)\right), \; J=\mathrm{det}(\mathbf{F}^M)\, .
\label{eq_invariants}
\eeq 

Furthermore, the invariants in eq. \ref{eq_invariants} require the definition of the isochoric left Cauchy-Green deformation tensor, 

\beq 
\bar{\mathbf{b}} = J^{-2/3}\mathbf{b} = J^{-2/3}\mathbf{F}^M(\mathbf{F}^{M} )^T \, .
\label{eq_bbar}
\eeq 

Note that the deformation metrics defined in eqs. (\ref{eq_invariants}) and (\ref{eq_bbar}) are with respect to the macroscale deformation. The strain energy function in (\ref{eq_Psi_M}) is not analytically available, but its existence implies that
  
\beq 
\mathbf{\sigma}^M = 2\mathbf{b} \frac{\partial \Psi}{\partial \mathbf{b}} = \mathbb{P}:\bar{\sigma} + p\mathbf{I} = \bar{\mathbf{\sigma}} - \frac{1}{3} \mathrm{tr}(\bar{\mathbf{\sigma}}) + p\mathbf{I} \, ,  
\label{eq_sigmaM}
\eeq 

where we have introduced the fictitious stress and the pressure,

\beq 
\bar{\mathbf{\sigma}} = 2\bar{\mathbf{b}}\frac{\partial \Psi_{\mathrm{iso}}}{\partial \bar{\mathbf{b}}}, \; p = \frac{\partial \Psi_{\mathrm{vol}}}{\partial J} \, .
\label{eq_sigmabar}
\eeq 

Finally, to compute the fictitious stress in (\ref{eq_sigmabar}), the derivatives $\Psi_1 (\bar{I}_1,\bar{I}_2) = \partial \Psi_{\mathrm{iso}}/\partial \bar{I}_1$ and $\Psi_2 (\bar{I}_1,\bar{I}_2) = \partial \Psi_{\mathrm{iso}}/\partial \bar{I}_2$ are needed, and, like the strain energy, these derivatives are also functions of the isochoric invariants. Thus, knowledge of the functions $\Psi_1, \Psi_2$, $\Psi_{\mathrm{vol}}$ are enough to compute the stress at the macroscopic level $\mathbf{\sigma}^M$. It is remarked again that these functions are not analytically available. The approach presented here relies on computing the stress and energy of the fibers from solving the RVE model with boundary condition $\mathbf{F}^M$. Then, a FCNN can be trained to represent the functions $\Psi_1, \Psi_2$ that best represent the stress data from the RVE simulations \textcolor{black}{without ever writing down an analytical expression for $\Psi$}.

\subsection*{Strain energy and its derivatives using Gaussian processes and optimization}

A Gaussian process (GP) is a stochastic process whose value at any collection of data points can be described with a multivariate normal distribution \cite{seeger2004gaussian}. Consider a data set comprised of input-output pairs $\mathcal{D}=\{\mathbf{x}=(I_1,I_2),\Psi_f\}$, where $I_1$ and $I_2$ are the invariants of the corresponding deformation $\mathbf{F}^M$ applied to the RVE, and $\Psi_f = \sum \Psi_f^{(i)}$ is the total strain energy accumulated in the fibers of the RVE. The data set $\mathcal{D}$  consists of the vector of inputs and outputs from all the training simulations. We used a GP to describe these data. The prior of the GP is fully determined by a mean function and a covariance function. A zero-mean prior was considered. For the covariance function we used a radial basis function (RBF) kernel \cite{scholkopf2004kernel}. This kernel is also commonly referred to as the squared exponential kernel. The implementation of the GP in the Python package \textit{GPy} was used \cite{sml2012sheffieldml}. The hyperparameters of this kernel are the length scale and signal strength, which were optimized to maximize the likelihood of the observed data pairs. After fitting, the posterior GP is denoted $\Psi_f^*(I_1,I_2)$, and it can be used to make predictions at points $\mathbf{x}^*=(I_1^*,I_2^*)$ that were not part of the training data. The predicted strain energy $\Psi_f^*$ at a new point $\mathbf{x}^*$ is a GP with mean function 

\beq 
\mu_f = \mathbf{k}(\mathbf{x}^*,\mathbf{x}) \mathbf{K}(\mathbf{x},\mathbf{x})^{-1} \Psi_f 
\label{eq_muf}
\eeq 

where $\mathbf{k}(\mathbf{x}^*,\mathbf{x})$ is the vector of covariances between the new test point $\mathbf{x}^*$ and the training data points $\mathbf{x}$. The matrix $\mathbf{K}(\mathbf{x},\mathbf{x})$ is the covariance between all training inputs. The vector $\Psi_f$ contains all the observations of the strain energy of the fibers for all the inputs $\mathbf{x}$. The posterior covariance is

\beq 
\Sigma_f = \mathbf{K}(\mathbf{x}^*, \mathbf{x}^*) - \mathbf{k}(\mathbf{x}^*,\mathbf{x}) \mathbf{K}(\mathbf{x},\mathbf{x})^{-1} \mathbf{k}(\mathbf{x}^*,\mathbf{x}) \, .
\label{eq_Sigf}
\eeq 

Given the posterior GP for the strain energy, $\Psi^*_f$, the derivatives of the strain energy function with respect to the invariants are also GPs that can be easily computed from eqs. (\ref{eq_muf}) and (\ref{eq_Sigf}) \cite{frankel2020tensor}. The mean of the gradient of the GP is simply the derivative of the mean function (\ref{eq_muf}), 

\beq 
\mu_{\Psi i} = \mathbf{k}_i(\mathbf{x}^*,\mathbf{x}) \mathbf{K}(\mathbf{x},\mathbf{x})^{-1} \Psi_f \, ,
\label{eq_muf}
\eeq 

where $\mathbf{k}_i(\mathbf{x}^*,\mathbf{x})$ is the vector with the derivatives of the kernel $\partial{k}((I_1^*,I_2^*,\mathbf{x})/\partial I^*_i$. The covariance of the derivatives can be obtained from the derivative of the kernel function,

\beq 
\Sigma_{\Psi i,\Psi j} = \frac{\partial^2 k(\mathbf{x},\mathbf{x}')}{\partial I_i \partial I'_j} \, .
\eeq 

However, the GPs for the derivative functions defined with eq. (\ref{eq_muf}) ignore the stress information and rely solely on the energy information. Thus, we furthered constrained the derivatives of the strain energy based on the stress data. In the previous section, the stress for a nearly incompressible material was derived.This description was the one that we eventually implemented in the finite element subroutine. In the RVE simulations incompressibility could be imposed exactly. For incompressible plane stress behavior the stress can be simpliefied to

\beq 
\mathbf{\sigma} = 2\mathbf{b}\frac{\partial \Psi_{\mathrm{iso}}}{\partial \mathbf{b}} + p\mathbf{I}
\label{eq_sigma_incompressible}
\eeq 

with $\mathrm{det}(\mathbf{b})=\mathrm{det}(\mathbf{F}^M)=1$, and where $p$ becomes a Lagrange multiplier that can be solved for to ensure the plane stress condition $\sigma_{zz}=0$. Expanding the derivative of the strain energy in (\ref{eq_sigma_incompressible}), 

\beq 
\mathbf{\sigma} = 2\Psi_1(I_1,I_2) \mathbf{b} +2 \Psi_2(I_1,I_2 )(I_1\mathbf{b} -\mathbf{b}^2) + p\mathbf{I} \, .
\label{eq_sigma_incompressible2}
\eeq 

To summarize, from the set of RVE simulations a GP was constructed over the total strain energy of the fibers as a function of the invariants of the applied deformation. This GP ignored the stress data initially. However, a relationship between the stress and the derivatives of the strain energy is available as seen in eq. (\ref{eq_sigma_incompressible2}). \textcolor{black}{Therefore, the stress calculated from eq. \ref{eq_sigma_micro} from each RVE was used to adjust the GP of the strain energy and its derivatives. In practice, this additional optimization step only changed the strain energy derivatives slightly, indicating that the strain energy and the stress were consistent with each other. Ultimately, the strain energy derivatives were used to train the neural network and the stress was no longer used at that stage. }The optimization was carried out with the \textit{minimize} function in the \textit{optimization} module of the Python package SciPy \cite{virtanen2020scipy}. At the end of this step we had all the data as illustrated in Figure \ref{fig02}A.  

\textcolor{black}{The approach described here differs from other data-driven approaches in the literature. For example, previous work using neural networks to describe material behavior has often focused on direct prediction of the stress components  \cite{chung2021neural,yang2020prediction}. The work by Vlassis et al. \cite{vlassis2020geometric}, on the other hand, focused on the prediction of the strain energy, similar to what is proposed here. However, in contrast to  \cite{vlassis2020geometric}, we retrieved the strain energy derivatives through the GP rather than just the energy. This step enabled training of the FCNN on the derivative functions, as described in the next section. The idea is similar to enhanced strain methods or other multi-field methods in elasticity where additional degrees of freedom are introduced and linked by additional constraint equations. The motivation is similar to multi-field problems in elasticity; predicting the energy derivatives as a direct output gives flexibility to the FCNN and results in smooth first and second derivative function predictions. In Vlassis et al. \cite{vlassis2020geometric}, regularization of the derivative predictions was done by using the Sobolev norm. Alternatively, Teichert et al. \cite{teichert2019machine} proposed integral neural networks that are trained on derivative information directly but do not require additional constraints to tie the derivatives to the energy function.}

\subsection*{Fully connected neural network}

 Based on the previous steps, schematized in Figure \ref{fig02}A, the overall input data consisted of  $\{ \lambda_x, \lambda_y, I_1, I_2, \theta, \varphi \}$, including multiple realizations of the networks even for the same $\{\theta, \varphi\}$. However, not all these input variables are independent. Ultimately, the goal was to obtain a model that was a function of the deformation invariants. Yet, it was easier to generate the training data by imposing the principal stretches. Thus, deformations on the RVE were imposed by sampling the in-plane stretches $\{ \lambda_x, \lambda_y \}$ which were post-processed to get the invariants $\{I_1, I_2\}$. Similarly, the output consisted of the variables $\{ \mathbf{\sigma},p,\Psi,\Psi_1,\Psi_2\} $. For a finite element implementation, only prediction of the functions $\{\Psi_1,\Psi_2\}$ was needed. Nevertheless, two FCNN were trained. A FCNN to predict the stress tensor directly as a function of the stretches $\lambda_x$, $\lambda_y$ was trained. However, this approach imposes the choice of a Cartesian coordinate system aligned with the principal in-plane directions for the strain and stress tensors \textcolor{black}{\cite{chung2021neural,yang2020prediction}}. Therefore this network was not explored in much detail.  An alternative approach, favored in this work, was to train the FCNN to predict the derivatives of the strain energy, $\Psi_1, \Psi_2$, as function of the strain invaraints. This second approach was the one implemented in the commercial finite element software Abaqus. In addition to the advantage of an invariant-based formulation, training the FCNN on the strain energy and its derivatives enabled the imposition of meaningful constraints, namely the convexity of the strain energy. \textcolor{black}{It also enabled the computation of the elasticity tensor with the second derivatives of the strain energy obtained through back-propagation. The elasticity tensor, in turn, was needed to solve nonlinear equilibrium problems, as described below. In contrast, previous work on data-driven constitutive models focused on predicting the energy and the stress, and paid less attention to the robust computation of the elasticity tensor needed for nonlinear equilibrium problems \cite{vlassis2020geometric}.} The process of FCNN training is illustrated in Figure \ref{fig02}B.

For the FCNN trained to predict $\Psi_1$ and $\Psi_2$, a weighted combination of the mean squared error (MSE) loss, mean absolute error (MAE) loss, and the mean absolute percentage error (MAPE) loss were used

\beq
\begin{aligned}
\mathbb{L}^{\mathrm{MSE}}_{i} &=& \frac{1}{N}\sum_{n=1}^{\# data} \left(\Psi_i^{(n)} - \hat{\Psi}_i^{(n)}\right)^2\\
\mathbb{L}^{\mathrm{MAE}}_{i} &=& \frac{1}{N} \sum_{n=1}^{\# data} \left|\Psi_i^{(n)} - \hat{\Psi}_i^{(n)}\right|\\
\mathbb{L}^{\mathrm{MAPE}}_{i} &=& \frac{100}{N}  \sum_{n=1}^{\# data} \left| \frac{\Psi_i^{(n)} - \hat{\Psi}_i^{(n)}}{\Psi_i^{(n)}}\right| \\
\end{aligned}
\label{eq_loss1}
\eeq 

where $i=1,2$, $n$ denotes each one of the $N$ data pairs  $\{(\varphi^{(n)},\theta^{(n)},I_1^{(n)},I_2^{(n)}),(\Psi_1^{(n)},\Psi_2^{(n))})\}$, and $\hat{\Psi}_i^{(n)}$ is the prediction of the strain energy derivative by the FCNN for the corresponding input  $(\varphi^{(n)},\theta^{(n)},I_1^{(n)},I_2^{(n)})$. The losses in (\ref{eq_loss1}) only penalize the error in the prediction of $\Psi_1$ and $\Psi_2$. However, as mentioned previously, an advantage of assuming a hyperelatic framework is that physically realistic constraints can also be considered during training of the FCNN. Indeed, a major driver for the development of analytical expressions for strain energy potentials of soft tissue is that convexity can be analyzed in detail \cite{chagnon2015hyperelastic}. 

\textcolor{black}{The requirement for $\Psi$ to have physically admissible solutions for a boundary value problem in nonlinear elasticity is that of polyconvexity with respect to $\mathbf{F}$ \cite{ball1976convexity,schroder2010anisotropic}. The condition of polyconvexity is, however, difficult to impose. An independent but related condition is convexity of the strain energy with respect to $\mathbf{C}$ \cite{lehmich2014convexity}. Under certain conditions, convexity of $\Psi$ with respect to $\mathbf{C}$ can be sufficient for the existence of global minimizers for boundary value problems in nonlinear elasticity \cite{gao2017convexity}. The condition reads \cite{gao2017convexity}}

\beq
\textcolor{black}{\Psi(\mathbf{C}) - \Psi(\mathbf{C}_0) \geq \left.\frac{\partial \Psi}{\partial \mathbf{C}}\right|_{\mathbf{C}_0} :(\mathbf{C}-\mathbf{C}_0), }
\label{eq_convexity_def}
\eeq 

\textcolor{black}{for all $\mathbf{C}\in \mathrm{Sym}^+$ and all admissible $\mathbf{C}_0$ (derived from an admissible deformation gradient). Alternatively, the convexity of $\Psi$ with respect to $\mathbf{C}$ is equivalent to the monotonicity of the second Piola Kirchhoff stress tensor with respect to changes in $\mathbf{C}$, i.e. }

\beq 
\textcolor{black}{
(\mathbf{S}_2-\mathbf{S}_1):(\mathbf{C}_2-\mathbf{C}_1)\geq 0 \, \forall \, \mathbf{C}_2, \mathbf{C}_1 \in \mathrm{Sym}^+ \, ,}
\label{eq_monotonicity}
\eeq 

\textcolor{black}{with $\mathbf{S}_i=2\partial \Psi /\partial \mathbf{C}_i$ the second Piola Kirchhoff stress corresponding to the deformation $\mathbf{C}_i$. We enforce the monotonicity of $\mathbf{S}$ with respect to $\mathbf{C}$ during training through an additional loss term that evaluates eq.  (\ref{eq_monotonicity}) for several pairs of $\mathbf{C}_1$, $\mathbf{C}_2$ across the input space, }

\beq 
\textcolor{black}{
\mathbb{L}^{\scas{convex}} = \sum_j \sum_i \max\left(-(\mathbf{S}_j-\mathbf{S}_i):(\mathbf{C}_j-\mathbf{C}_i),0\right) } \, .
\label{eq_loss_monotonicity}
\eeq 

\textcolor{black}{Additionally, consider the Hessian of the strain energy, }

\beq
\mathbf{H}^{(n)} =  \left(\begin{array}{cc}
                \Psi_{11} & \Psi_{12}\\
                \Psi_{21} &\Psi_{22}
                \end{array} \right) \, =
                \left(\begin{array}{cc}
                \frac{\partial \hat{\Psi}^{(n)}_1}{\partial I_1} & \frac{\partial \hat{\Psi}^{(n)}_1}{\partial I_2}\\
                \frac{\partial \hat{\Psi}^{(n)}_2}{\partial I_1} & \frac{\partial \hat{\Psi}^{(n)}_2}{\partial I_2}
                \end{array} \right) \, .
\label{eq_hessian}
\eeq 

The Hessian contains the second derivatives of the strain energy with respect to the invariants. On the other hand, since the FCNN output are already the first derivatives $\Psi_1, \Psi_2$,  only one derivative of the FCNN output is required. Computing the derivatives of the FCNN output with respect to the inputs is straightforward and follows an algorithm analogous to back-propagation \cite{avrutskiy2017backpropagation}. The first obvious constraint on the FCNN output is that the Hessian must be \textcolor{black}{symmetric}. Therefore, one additional component of the loss considered here is

\beq 
\mathbb{L}^{\mathrm{sym}} = \sum_{n=1}^{\# data} \left| \frac{\partial \hat{\Psi}^{(n)}_1}{\partial I_2} - \frac{\partial \hat{\Psi}^{(n)}_2}{\partial I_1} \right|
\eeq 

\textcolor{black}{Under certain conditions, positive semi-definiteness of $\mathbf{H}$ can be used to enforce convexity with respect to $\mathbf{C}$ and even polyconvexity with respect to $\mathbf{F}$ \cite{itskov2006polyconvex,schroder2003invariant}. The advantage of this approach is that it requires only local calculations, as opposed to eq.  (\ref{eq_loss_monotonicity}), which couples the entire input space. However, this additional constraint on the loss was not considered in the main text because it is not necessary to satisfy convexity with respect to $\mathbf{C}$. In fact, positive definiteness of $\mathbf{H}$ is too restrictive, which is discussed further in the Supplemental Material.  } 


The final loss function  can be obtained by adding the different losses, each with a given weight. The weights were manually adjusted to obtain desired accuracy and satisfaction of the constraints.

An additional FCNN for the prediction of the principal stresses as a function of the principal stretches was also constructed. However, since in the end this FCNN was not implemented into the finite element code, only the standard MAPE loss was used in training, which is a standard in machine learning \cite{botchkarev2018performance,bowerman2005forecasting}. 

The Adam optimization algorithm \cite{kingma2014adam} was used to train the networks and update the FCNN weights during back-propagation \cite{rumelhart1986learning}. The initial learning rate was set as $0.0001$ with a momentum decay of $0.9$. The training and validation data splits were $85$\% and $15$\% of the total data points. The total number of observations was $132,000$ for the $1,100$ networks and the different deformations.  A typical FCNN consists of an input layer, one or more hidden layers, and an output layer. Here we trained and evaluated an FCNN with 3 hidden layers of dimensions 8, 8, and 16 respectively, and 1 output layer. The activation function used was ReLu \cite{nair2010rectified}(Table \ref{table01}).
The training was implemented using Keras \cite{chollet2018keras} with a TensorFlow \cite{abadi2016tensorflow} back-end on a hardware platform with the following specifications: Overclocking 5.0 GHz Intel i9 processor, 32 GB DDR4/2666 MHz memory, and Nvidia GeForce GTX 1080. The batch size, which controls the number of samples to be propagated through the network at a time was set as 64 to avoid local minima, enhance generalization performance, and improve the optimization convergence \cite{keskar2016large}. 

\begin{table}[h!]\centering
\caption{NN architectures}
\label{table01}
\begin{tabularx}{0.45\textwidth}{lll}
\hline
Layer (type)&Output shape&\# Parameters\\ \hline
Dense 1&8&40\\ 
Activation 1&8&0\\ 
Dense 2&8&72\\
Activation 2&8&0\\
Dense 3&16&144\\
Activation 3&16&0\\
Dense 4&2&34\\
Activation 4&2&0\\ \hline
\end{tabularx}
\end{table}

\subsubsection*{Finite element implementation}

A general FCNN-based user material subroutine (UMAT) was implemented in the nonlinear finite element package Abaqus Standard (Rhode Island, United States). Rather than hard-coding the specific FCNN that was trained on the DFN data, the UMAT function was programmed to take in the network architecture and the weights and biases as material parameters. Thus, the FCNN needs to be specified in the input file. As a result, the UMAT can be used for different FCNN architectures and trained on different data. In the UMAT function itself, the FCNN is constructed and evaluated based on the information of the input file. This approach gives flexibility. For example, in this paper two different regions in a finite element model were assigned different FCNNs to model their material behavior without the need for multiple user subroutines. The deformation gradient $\mathbf{F}$ at an integration point is passed as one of the arguments of the UMAT. Based on this deformation gradient, the corresponding isochoric left Cauchy Green deformation, $\bar{\mathbf{b}}$,  and its invariants $\bar{I}_1, \bar{I}_2$, are computed. Then, using the microstrucure parameters $\varphi$ and $\theta$ which are defined as material properties, together with the network architecture and the weights and biases of the network (also defined as material properties), the FCNN evaluates $\Psi_1, \Psi_2$. Evaluation of neural networks is a combination of  simple operations. Consider a dense layer, layer $i$, of the network. Layer $i$ takes in $m$ inputs from the previous layer stored in vector $\mathbf{y}_{i-1}$ and produces $n$ outputs $\mathbf{y}_i$. In other words, $m$ is the number of neurons in layer $i-1$ and $n$ is the number of neurons in the layer $i$. The weights for this layer are arranged in the matrix $\mathbf{W}_i$ which is $m \times n$. The vector of biases for layer $i$ is denoted $\mathbf{B}_i$. Propagation of information in this layer corresponds to the linear map 

\beq 
\mathbf{y}_{i} = \mathbf{W}_{i}^T \mathbf{y}_{i-1} + \mathbf{B}_i \, .
\eeq 

The nonlinearity in the network comes from the activation layers. For the activation layer $i$, the output is simply

\beq 
\mathbf{y}_{i} = g( \mathbf{y}_{i-1} )
\eeq 

where $g(\bullet)$ is the activation function applied element-wise to the vector $\mathbf{y}_{i-1}$. The ReLu activation function used here is $g(y) = max(0,y)$. 
In addition to computing the functions $\Psi_1, \Psi_2$ , their derivatives with respect to the invariants $\bar{I}_1, \bar{I}_2$ are needed for the elasticity tensor, which is essential for the convergence of the Newton iterations of the finite element solver. For a dense layer $i$, the input gradient is $J_{i-1}$ of dimension $m \times n_{\mathrm{input}} $, with $n_{\mathrm{input}}$  the number of inputs to the whole FCNN. The gradient output of a dense layer $i$ is  

\beq 
\mathbf{J}_i = \mathbf{W}^T_{i} \mathbf{J}_{i-1} \, . 
\label{eq_J1}
\eeq 

For the activation layer, the gradient output is 

\beq 
\mathbf{J}_i = \mathrm{diag}(g'(\mathbf{y}_i)) \mathbf{J}_{i-1} \, .
\label{eq_J2}
\eeq 

where $\mathrm{diag}(\bullet)$ denotes a diagonal matrix, and $g'(\bullet)$ is the derivative of the activation function. The gradient for the input layer is initialized as the identity matrix of size $n_{\mathrm{input}} \times n_{\mathrm{input}} $. After doing the operations defined in (\ref{eq_J1}) and (\ref{eq_J2}), the gradient output is a matrix of size $n_{\mathrm{output}}\times n_{\mathrm{input}}$. For the particular FCNN used in this paper, the final gradient output contains the Hessian (\ref{eq_hessian}) as a sub-matrix. 

After evaluation of the FCNN, the prediction of $\Psi_1$ and $\Psi_2$ are used to compute the isochoric stress based on equations (\ref{eq_sigmaM}) and (\ref{eq_sigmabar}). Additionally, a volumetric strain energy is required, which is not part of the FCNN prediction. Here we opt for a classical expression for the volumetric part of the strain energy

\begin{equation}
\Psi_{\mathrm{vol}} = \frac{K}{2}(J-1)^2
\end{equation}

where $K$ is the bulk modulus and $J=\mathrm{det}(\mathbf{F})$ is the volume change. 

The consistent tangent can be derived independently of the material model, simply by knowing the second derivatives of the strain energy function. The FCNN outputs the prediction of the second derivatives based on the operations (\ref{eq_J1}) and (\ref{eq_J2}) . The following terms are used for ease of notation of the consistent tangent  

\beq 
\begin{split}
\delta_1 &= 4(\Psi_2 + \Psi_{11} + 2\bar{I}_1^2 \Psi_{22})\\
\delta_2 &= -4(\Psi_{12} + \bar{I}_1 \Psi_{22})\\
\delta_3 &= 4\Psi_{22}\\
\delta_4 &= -4 \Psi_2 
\end{split}
\eeq 

where $\Psi_{11}, \Psi_{12}, \Psi_{21}, \Psi_{22}$ are the derivatives predicted by the FCNN that correspond to the entries of the Hessian, eq. (\ref{eq_hessian}). The isochoric part of the elasticity tensor is

\beq 
\begin{split}
J \mathbb{c}_{\mathrm{iso}} =&  \delta_1 \left(\mathbf{\bar{b}} \otimes \mathbf{\bar{b}} - \frac{1}{3} \bar{I}_1(\mathbf{\bar{b}} \otimes \mathbf{I} + \mathbf{I}\otimes \mathbf{\bar{b}}) + \frac{1}{9} \bar{I}_1^2 \mathbf{I}\otimes \mathbf{I} \right)\\
&+\delta_2 \left(\mathbf{\hat{b}} \otimes \mathbf{\bar{b}}^2 + \mathbf{\bar{b}}^2 \otimes \mathbf{\bar{b}} -\frac{1}{3} \mathrm{tr}(\mathbf{\bar{b}}^2)(\mathbf{\bar{b}}\otimes \mathbf{I}+\mathbf{I}\otimes \mathbf{\bar{b}})-\frac{1}{3}\bar{I}_1(\mathbf{\bar{b}}^2\otimes\mathbf{I}+\mathbf{I}\otimes \mathbf{\bar{b}}^2)+\frac{2}{9} \bar{I}_1 \mathrm{tr}(\mathbf{\bar{b}}^2)\mathbf{I}\otimes \mathbf{I}\right)\\
&+\delta_3 \left(\mathbf{\bar{b}}^2\otimes \mathbf{\bar{b}}^2 - \frac{1}{3}\mathrm{tr}( \mathbf{\bar{b}}^2)(\mathbf{\bar{b}}^2\otimes \mathbf{I} + \mathbf{I}\otimes \mathbf{\bar{b}}^2)+\frac{1}{9}\left(\mathrm{tr}(\mathbf{\bar{b}}^2)\right)^2\mathbf{I}\otimes \mathbf{I}\right)\\
&+\delta_4 \left(\mathbf{\bar{b}}\bar{\otimes} \mathbf{\bar{b}} -\frac{1}{3}(\mathbf{\bar{b}}^2\otimes \mathbf{I}+\mathbf{I}\otimes \mathbf{\bar{b}}) +\frac{1}{9}(\mathbf{\bar{b}}:\mathbf{\bar{b}})\mathbf{I}\otimes\mathbf{I}\right)
\end{split}
\label{eq_cciso}
\eeq 

where the following special dyadic product was used $(\bullet)_{ij} \bar{\otimes} (\circ)_{kl} = (\star)_{ikjl}$ . The volumetric contribution of the elasticity tensor is

\beq 
\mathbb{c}_{\mathrm{vol}} = \left(p+J\frac{\partial p}{\partial J}\right)\mathbf{I}\otimes \mathbf{I} -2p\mathbf{I} \bar{\otimes} \mathbf{I} \, .
\label{eq_ccvol}
\eeq 

The isochoric and volumetric components of the spatial elasticity tensor defined in eqs. (\ref{eq_cciso}) and (\ref{eq_ccvol}) are part of the consistent tangent needed for Abaqus' Newton-Raphson solver. Mechanical equilibrium is solved in Abaqus by incrementally integrating the stress using the Jaumann stress rate. As a result, the consistent tangent needs a correction with respect to the standard form of the elasticity tensor \cite{holzapfel2000nonlinear}. The tangent for Abaqus takes the form 

\begin{equation}
\mathbb{c}_{\mathrm{abaqus}} = \mathbb{c}_{\mathrm{iso}} + \mathbb{c}_{\mathrm{vol}} + \frac{1}{2}(\mathbf{\sigma} \overline{\otimes} \mathbf{I} + \mathbf{\sigma} \underline{\otimes} \mathbf{I} + \mathbf{I} \overline{\otimes} \mathbf{\sigma} + \mathbf{I} \underline{\otimes} \mathbf{\sigma})
\end{equation}

in which an additional tensor product is defined as $(\bullet)_{ij} \bar{\otimes} (\circ)_{kl} = (\star)_{iljk}$. In summary, the architecture, weights, and biases of the FCNN are passed to the UMAT subroutine as material parameters, and the FCNN is evaluated given the current deformation gradient at an integration point to output $\Psi_1,\Psi_2$ and their derivatives $\Psi_{11},\Psi_{12}, \Psi_{21},\Psi_{22}$. These functions are used to compute the stress $\mathbf{\sigma}$ and the consistent tangent $\mathbb{c}_{\mathrm{abaqus}}$ .

\section*{Results}

\subsection*{Response under uniaxial and biaxial loading}

We first compared that the results from our DFN implementation matched observations from previous computational models of fiber networks as well as experimental data on fibrin and collagen gels. We created 110 networks with a volume fraction $\theta \approx 0.3 \%,\, \theta \in (0.297\%,0.303\%) $ and fiber diameter $\varphi=100$nm. The volume fraction could not be controlled exactly because the networks were generated by seeding the RVE with randomly located nucleation points and letting them grow in random directions until the algorithm was terminated. Due to these sources of randomness, control of the exact fiber length was not possible. Nevertheless, the range of volume fractions obtained was very close to the desired value of $0.3\%$. We created multiple networks with the same values $\theta$ and $\varphi$ because these variables describe the homogenized network properties but do not specify a network uniquely. 

\begin{figure}
\centering
\includegraphics[width=0.98\linewidth]{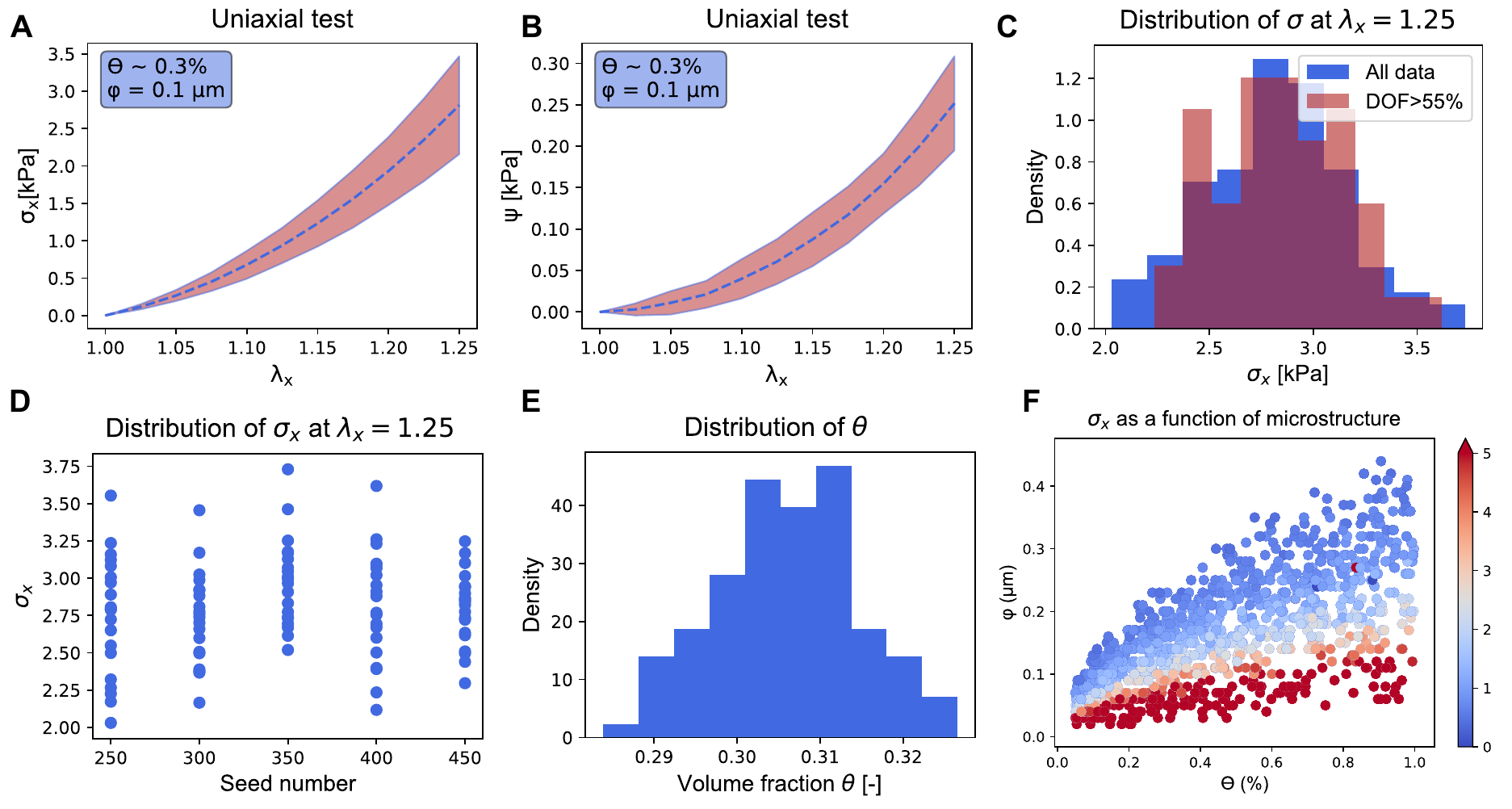}
\caption{Verification of DFN response under uniaxial deformation. In uniaxial tests, 110 networks with homogenized microstructure properties $\theta\approx 0.3\%$, $\varphi=100$nm showed nonlinear behavior for the stress (A) and the strain energy density (B). Even though the networks approximately shared the same homogenized properties $\theta$, $\varphi$, the random nature of the networks led to uncertainty, shown as a shaded area in (A), \textcolor{black}{and (B) for the stress and strain energy respectively. The variance in the stress $\sigma_x$  was independent on the number of degrees of freedom of the network, defined as the percentage of inner nodes with respect to total number of nodes (C). Another verification was to plot the stress as a function of the seed number, which also showed that the spread of the stress was not affected by the seed number (D). Thus, the variance in the stress was due to the inherent randomness of the networks and due to variation in the volume fraction $\theta$ which was not controlled exactly (E).  The stress in uniaxial tests was a function of the homogenized microstructure properties $\varphi$ and $\theta$ for a fixed value of fiber stiffness, with increasing stress for smaller fiber diameters and relatively little influence of the volume fraction (F)}.}
\label{fig03} 
\end{figure}

The 110 RVEs were subjected to uniaxial deformation following similar approaches in the literature \cite{lin2015influence,lai2012mechanical,stylianopoulos2007volume}. 
The nonlinear relationship of stress as a function of stretch was observed, with increasing slope at higher stretch values. At a stretch of $\lambda_x=1.2$, for example, the mean (+- S.D.) stress was 1.45 kPa +- 0.33 kPa (Figure \ref{fig03}A). \textcolor{black}{The strain energy density followed a similar trend (Figure \ref{fig03}B)}. The stress and strain energy showed variance, suggesting that even networks with the same homogenized microstructure properties $\varphi,\theta$, can have slightly different mechanical behavior. \textcolor{black}{To further explore this uncertainty, we plotted the variance in the stress at the end of the uniaxial loading (Figure \ref{fig03}C). The stress distribution for a subset of the data with a higher number of degrees of freedom, i.e. higher number of inner nodes with respect to the total number of nodes, was also visualized (Figure \ref{fig03}C). The agreement in the two histograms provided confidence that the simulations converged with respect to RVE size. We also plotted the stress at the end of uniaxial loading with respect to the seed number and observed that the spread stayed consistent as the number of seeds increased (Figure \ref{fig03}D). Therefore, variance in the stress and strain energy was due to the inherent randomness of the networks as well as variance in the volume fraction $\theta$ (Figure \ref{fig03}E). } The microstructure variables $\theta,\varphi$, led to consistent trends in the stress response under uniaxial deformation: decreasing fiber diameter led to increase in stress, almost independently of volume fraction (Figure \ref{fig03}C), \textcolor{black}{implying an increase in the number of fibers. Since the stiffness of the fibers was kept constant regardless of fiber diameter ($k_f=0.02$kPa), the increase in the number fibers is what led to stiffer RVEs}.

\begin{figure}
\centering
\includegraphics[width=0.98\linewidth]{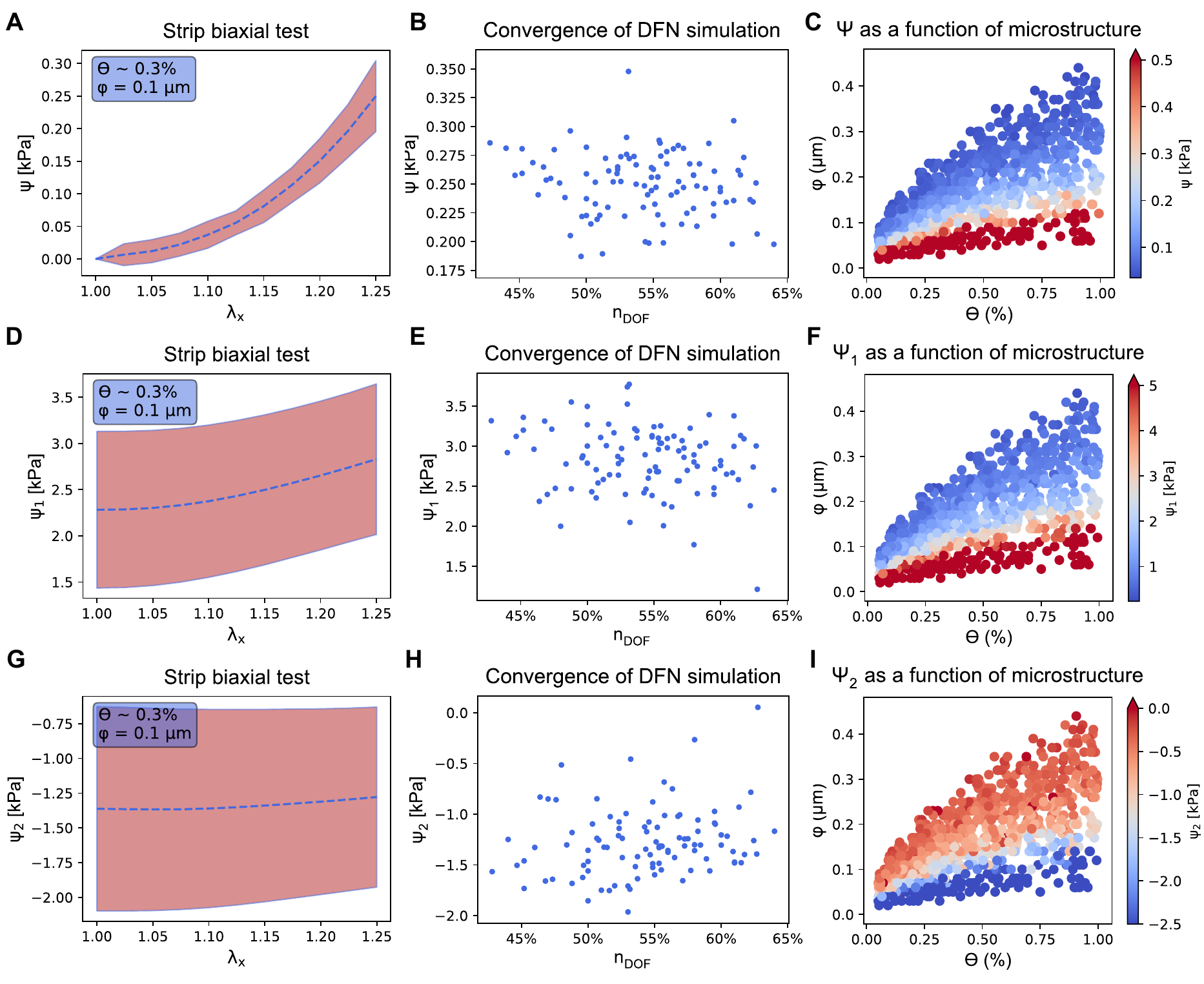}
\caption{Response of DFNs under strip-biaxial loading. The strain energy $\Psi$ (A), and its derivatives $\Psi_1,\, \Psi_2$ show a slightly nonlinear response with respect to increasing deformation (D, G). The strain energy variance depends on the number of degrees of freedom (B), while the derivatives are relatively constant even as the number of degrees of freedom increases (E, H). In response of changes in microstructure, the strain energy in the fibers increases with both volume fraction $\theta$  and fiber diameter $\varphi$ (C), while the derivatives $\Psi_1$ and $\Psi_2$ are mostly insensitive to the volume fraction but increase with fiber diameter (F, I). }
\label{fig04} 
\end{figure}

The response of the networks under biaxial deformations was also explored. Figure \ref{fig04} shows the response under strip-biaxial testing keeping $\lambda_y=1$. For the 110 networks with $\theta=0.3\%, \varphi=100$nm, the strain energy density in the fibers, $\Psi$, showed a similar trend to the uniaxial tests. The strain energy increased nonlinearly for increasing stretch, with some variance attributed to the randomness of the 110 networks even though they shared the same homogenized properties (Figure \ref{fig04}A).  We also explored the effect of the number of degrees of freedom on the variance of the strain energy. No significant variance of $\Psi$ could be found as the number of degrees of freedom increased (Figure \ref{fig04}B). Analyzing the mechanical behavior with respect to the microstructure variables, it can be seen that low volume fraction and smaller diameters resulted in higher energy density (Figure \ref{fig04}B). The derivatives, $\Psi_1$ and $\Psi_2$, showed a slightly nonlinear response with a relatively constant variance with respect to the deformation. The variance in the derivatives did not show a dependence on the number of degrees of freedom (Figure \ref{fig04}E,H). Analyzing the contours of the functions $\Psi_1$ and $\Psi_2$, it can be seen that \textcolor{black}{$\Psi_1$ depends on the microstructure variables $\theta$ and $\varphi$ in a similar way to the stress (Figure \ref{fig04}F,I), while $\Psi_2$ showed an inverse relationship, with decreasing values for smaller fiber diameters. On the other hand, the absolute value of $\Psi_2$ increased analogously to $\Psi_1$ as a function of microstructure. The similarities in the contour trends between stress and strain energy derivatives was expected since the derivatives appear linearly in the definition of the stress tensor, see eq. (\ref{eq_sigma_incompressible2})}. Note that for a neo-Hookean material, the vaue of $\Psi_1$ is a constant and the value of $\Psi_2$ is zero. For a Mooney-Rivlin  material, the value of both $\Psi_1$ and $\Psi_2$ are generally non-zero constants. Our results show that the DFN networks are sightly more nonlinear that these simple but commonly used strain energy functions. Instead of trying to find an analytical expression with a higher degree of nonlinearity to match the data, the next section explains how the FCNN was trained to predict the mechanical response without the need of classical strain energy functions.

\subsection*{FCNN training and validation}

\begin{figure}
\centering
\includegraphics[width=0.9\linewidth]{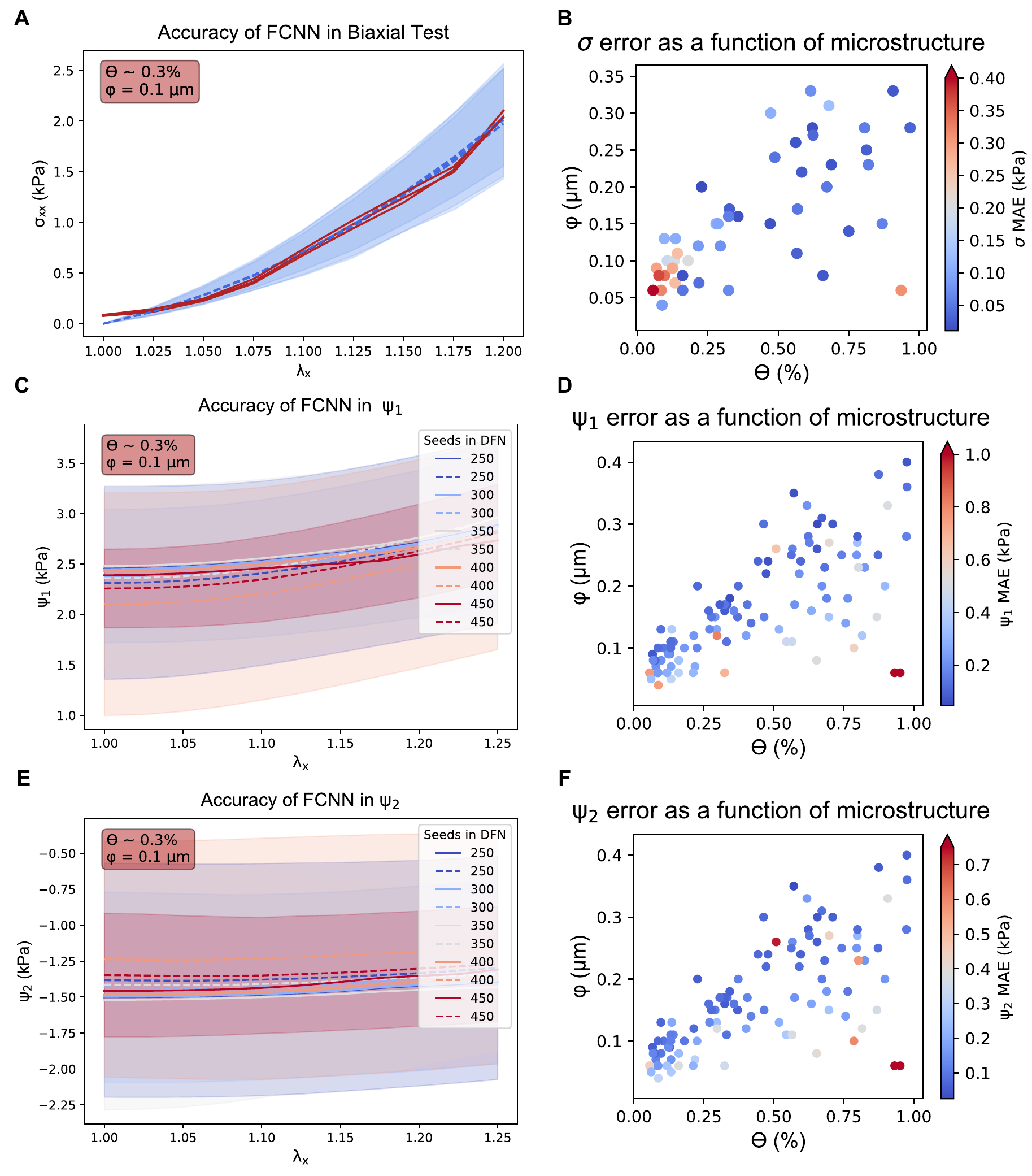}
\caption{Performance of the FCNN against the testing set. Performance of a neural network for the prediction of stress in strip biaxial loading  (solid red line) for showed agreement with the mean (dashed blue line) and confidence interval (shaded blue region) of 60 DFNs with the equivalent microstructure  microstructure $\theta=0.3\%, \varphi=100$nm (A). Error as a function of microstructure showed less accuracy of the FCNN for smaller volume fractions and fiber diameters (B). The derivatives of the strain energy predicted by the FCNN (solid red lines) also agreed with the means (dashed lines) and confidence intervals (shaded regions) of DFNs with equivalent microstructure (C,E). \textcolor{black}{The variance in the ground truth was not affected by the number of fibers in the network}. Loss function values for $\Psi_1$ and $\Psi_2$ were small over the microstructure range (D,F). \textcolor{black}{Absolute errors were higher near the boundaries of the region for which there is less training data for the FCNN, as well as for smaller fiber diameters. These trends align with the values of the stress and strain energy, which also increased in magnitude for smaller fiber diameters. } }
\label{fig05} 
\end{figure}

A total of \textcolor{black}{132,000} data points were used to train the FCNN with the architecture specified in Table \ref{table01}.  As seen in the uniaxial and biaxial response tests (Figures \ref{fig03} and \ref{fig04}), even for the same microstructure variables $\theta$ and $\varphi$, there is variance in the stress, the strain energy, and the derivatives of the strain energy. The FCNN trained on stress data was first compared to strip-biaxial data from 60 DFNs with the equivalent microstructure $\theta=0.3\%, \varphi=100$nm. Figure \ref{fig05}A shows the comparison of the FCNN prediction of the stress against the mean response of the 60 DFNs. The microstructure variables for this case, $\theta=0.3\%, \varphi=100$nm, were not part of the training of this FCNN. We found that the FCNN predictions (red solid line) laid well within the confidence interval (blue shading) and were very close to the mean value of the simulations of the 60 DFNs (blue dashed line in Figure \ref{fig05}A). During training, the raw data from each DFN was used, and not the mean response of different DFNs. The loss functions were evaluated with respect to the raw DFN data, i.e. without taking into account the variance in the DFN response. Thus, the FCNN converged naturally toward the mean response even though it was not trained directly on the mean data from multiple RVEs with the same $\theta$, $\varphi$. To explore the error as a function of microstructure, additional testing data consisting of \textcolor{black}{1000} simulations with different combinations of $\theta$ and $\varphi$ were evaluated under strip-biaxial loading. The results are shown in Figure \ref{fig05}B. In this case, the prediction of the FCNN was compared against  at least 20 fiber networks sharing equivalent $\theta, \varphi$ pairs that were not a part of the training data. \textcolor{black}{Errors were small in the majority of the validation cases, with slightly higher errors for small fiber diameters and volume fractions, near the boundaries of the training region. }

\begin{figure}
\centering
\includegraphics[width=0.9\linewidth]{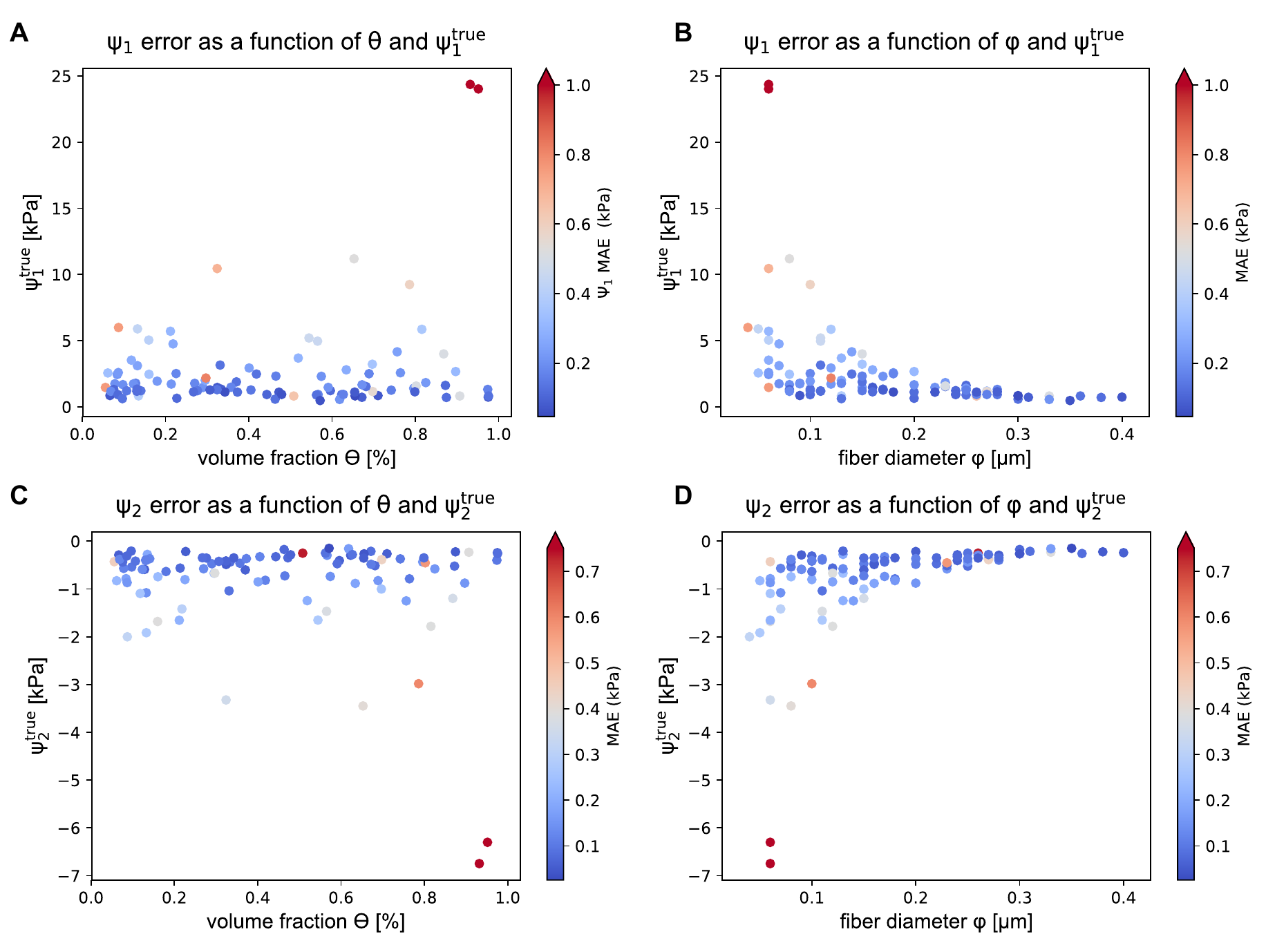}
\caption{Performance of the FCNN as a function of microstructure. Predictions of the derivative function $\Psi_1$ with respect to the volume fraction $\theta$ (A) and the fiber diameter $\varphi$ (B). The error of the derivative prediction $\Psi_2$ showed similar trends with respect to $\theta$ and $\varphi$ (C,D). \textcolor{black}{The validation data showed outlier values for which the FCNN prediction errors were large. However, errors were small in most of the validation set. }}
\label{fig06} 
\end{figure}

The FCNN that was trained directly on the derivative data $\Psi_1$ and $\Psi_2$ without convexity constraints was first compared to 100 DFNs with the equivalent microstructure $\theta=0.3\%, \varphi=100$nm, subjected to strip-biaxial loading. The DFNs in the validation set were separated based on the number of seeds used to initialize the networks. \textcolor{black}{The number of seeds did not affect the results}. The FCNN prediction was inside of the confidence intervals (shaded areas) of the DFNs (Figure \ref{fig05}C,E). As a function of the microstructure variables $\theta$ and $\varphi$, the model predicted $\Psi_1$, $\Psi_2$ with small errors for most networks (Figure \ref{fig05}D,F). \textcolor{black}{Higher absolute errors were observed for smaller fiber diameters, which aligns with the observation of higher absolute value of the stress and strain energy for decreasing fiber diameters.}


Figure \ref{fig06} further shows the dependence of the error on the microstructure while also showing the dependence on the true value of the derivative. Higher errors were seen on the edge of the plot, namely, for those networks with extreme volume fractions and fiber diameters, which showed extreme values of $\Psi$. While most of the values for $\Psi_1,\Psi_2$ were  concentrated within a narrow range, some outlier values were noticed in the validation data (Figure \ref{fig06}). As expected, the FCNN metamodel predictions were better on the region of the output space closer to the center of the range spanned by the true $\Psi_1$ and $\Psi_2$.

\subsection*{Comparison against conventional material models}

\textcolor{black}{Conventional material models can represent nonlinear behavior and have been used to describe biopolymer gels. Neo-Hookean and Ogden strain energies are the most common \cite{sugerman2021whole}. Comparison between the FCNN, neo-Hookean, and Ogden models for uniaxial, strip-biaxial, and equi-biaxial tests are shown in Figure \ref{fig_ogden_fit}. The neo-Hookean model is unable to capture the response of the RVEs. The Ogden model closely matches the microscale simulations. The FCNN performs similarly to the Ogden model. An important difference between the analytical strain energies and the FCNN is that the analytical models were fitted directly to the data shown in Figure \ref{fig_ogden_fit}, whereas the FCNN was trained on a broader data set in which both microstructure parameters ($\theta$, $\varphi$) were varied, and was trained based on the energy and the energy derivatives but not directly on the stress data. Thus, even though the FCNN was not trained directly on the data shown in Figure \ref{fig_ogden_fit} it was still able to predict accurately the response of these gels.} 

\setlength{\unitlength}{1cm}
\begin{figure}
\centering
\begin{picture}(15,6)
\put(0,0){\includegraphics[width=0.85\linewidth]{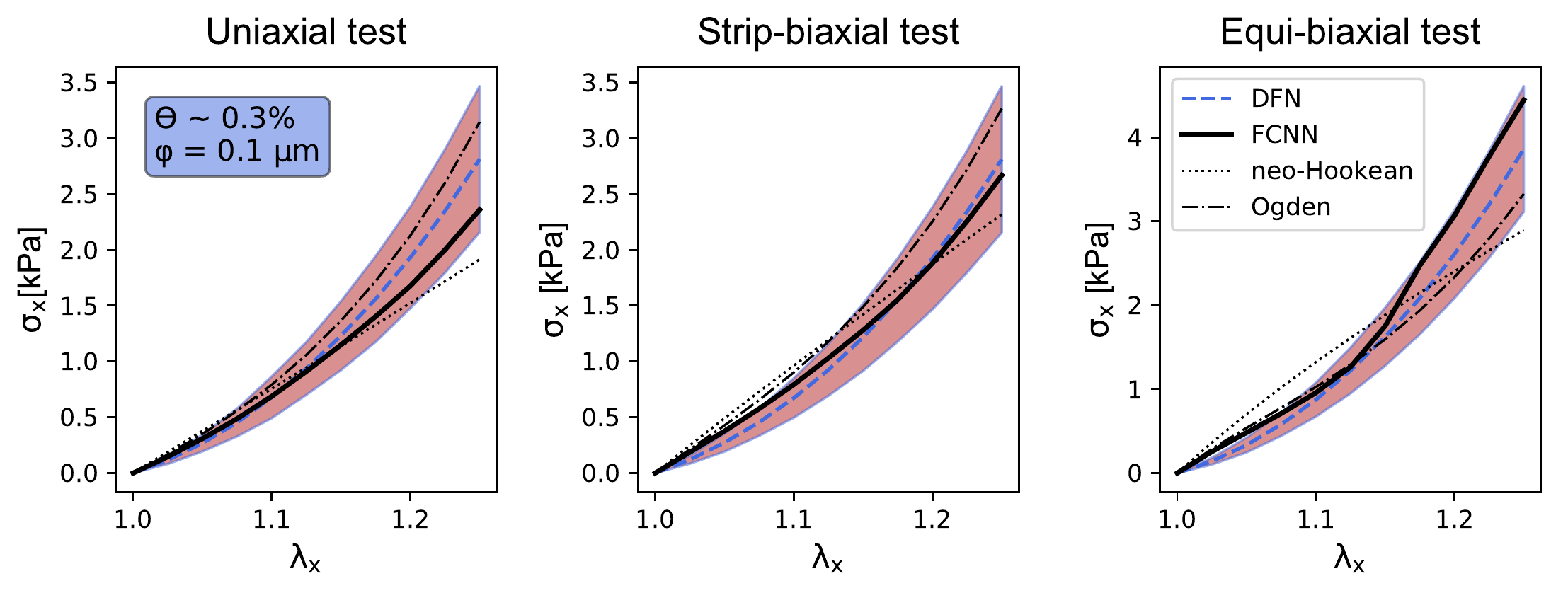} }
\end{picture}
\caption{\textcolor{black}{Comparison between the FCNN, neo-Hookean, and Ogden models against the DFN data for RVEs with microstructure $\theta~0.3$, $\varphi=0.1$ for uniaxial (A), strip-biaxial (B), and equi-biaxial (C) tests. The neo-Hookean and Ogden models were fitted directly to the data shown in the Figure whereas the FCNN was trained on the data set consisting of 1100 fiber networks in which both $\theta$ and $\varphi$ were varied. Additionally, the FCNN was trained on the energy derivatives and not on the stress data directly.} }
\label{fig_ogden_fit} 
\end{figure}

\subsection*{Convexity of the FCNN model}

\textcolor{black}{We first performed convexity checks on the DFN data used to train the neural network. For ten of the RVEs, all $120\times 120$ possible comparisons between pairs of input points were used to evaluate eqs. (\ref{eq_convexity_def}) and (\ref{eq_monotonicity}). The results are summarized in Table \ref{table02}. These results show that the data used to train the FCNN was already representative of an underlying convex strain energy function. The convexity tests were also evaluated for the FCNN without convexity constraints. Ten different microstructures were sampled and their response evaluated using the FCNN. Then, pair-wise comparisons between the outputs were done according to eqs. (\ref{eq_convexity_def}) and (\ref{eq_monotonicity}), see Table \ref{table02}. Due to the underlying convexity in the data, the FCNN predictions satisfied the convexity requirement even without imposing the constraints during training.  When the strain energy contours for the microstructure $\theta=0.3$, $\varphi=0.1\mu\mathrm{m}$ were plotted, it can be seen that the strain energy is convex with respect to the Green Lagrange strain, which is equivalent to convexity with respect to $\mathbf{C}$ (Figure \ref{fig07}). To generate Figure \ref{fig07}, derivatives $\Psi_1, \Psi_2$ predicted by the FCNN were used to determine the second Piola Kirchhoff stress and integrated over strain curves $\int \mathbf{S}:d\mathbf{E}$.}

\setlength{\unitlength}{1cm}
\begin{figure}
\centering
\begin{picture}(15,6)
\put(0,0){\includegraphics[width=0.85\linewidth]{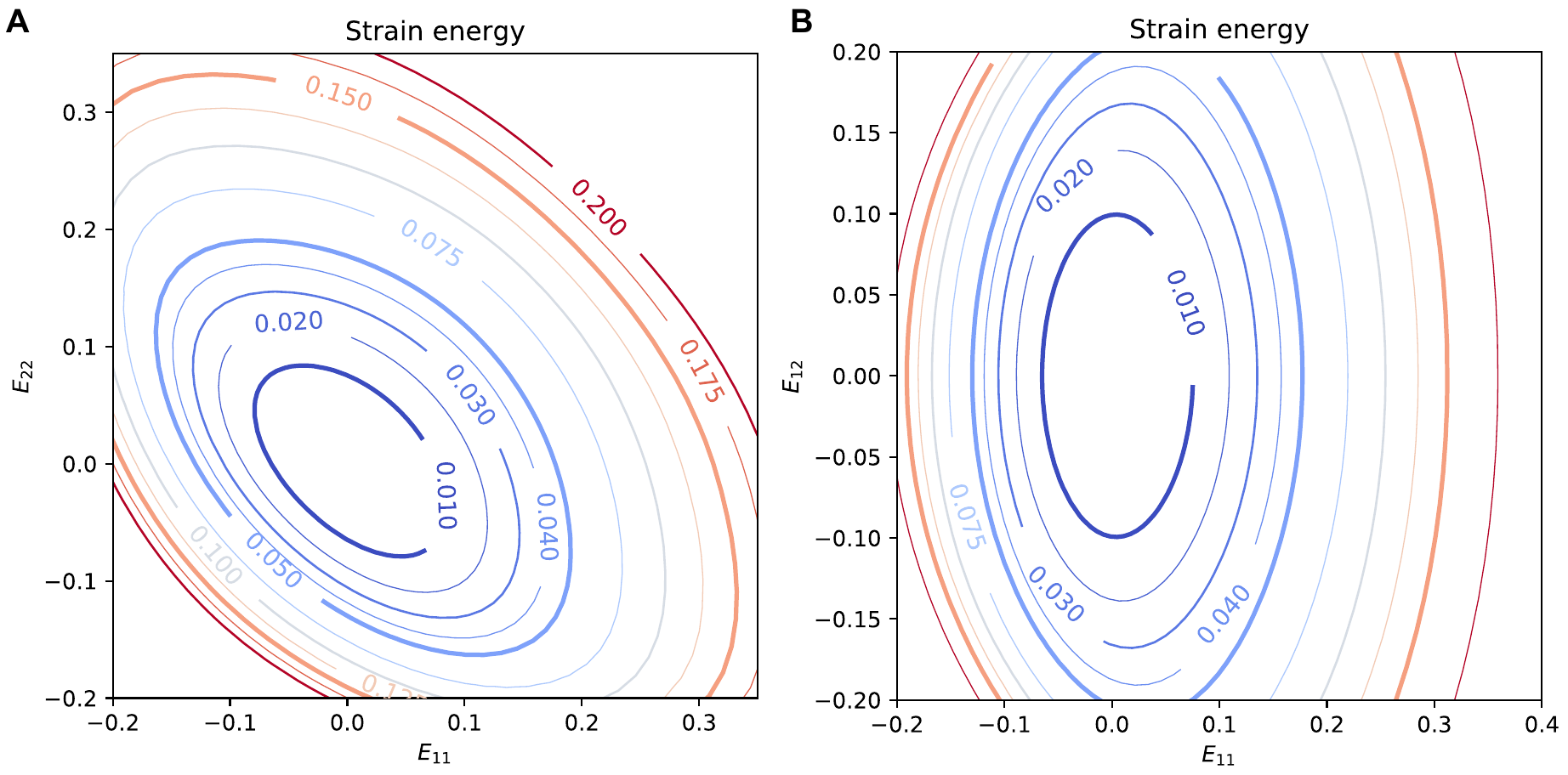} }
\end{picture}
\caption{Strain energy density function with respect to $E_{11}$ and $E_{22}$ with $E_{12} = 0$ for a network with microstructure ($\theta=0.3\%, \varphi=100$nm) (A), and with respect to $E_{11}$ and $E_{12}$ with $E_{22} = 0$ (B). }
\label{fig07} 
\end{figure}

\begin{table}[h!]\centering
\caption{Summary of convexity checks.}
\label{table02}
\begin{tabularx}{0.855\textwidth}{||l||c|c||}
\hline
&Pair-wise comparisons&Pair-wise comparisons\\
&satisfying eq. (17)&satisfying eq. (18)\\
\hline
RVE data &96.01\% &100\% \\
\hline 
FCNN without constraints &99.87\% &100\% \\
\hline 
FCNN trained with monotonicity constraint&98.99\% &99.83 \%\\
FCNN trained with invariant convexity&99.95\% &100 \%\\
\hline 
\hline 
Non-convex dataset&60.81\% & 63.4\%\\
\hline 
FCNN trained without constraints& 66.18\% & 73.6\%\\
\hline 
FCNN trained with monotonicity constraint&61.79\% &100 \%\\
\hline 
\end{tabularx}
\end{table}

\textcolor{black}{Training a FCNN while imposing the monotonicity of the second Piola Kirchhoff stress tensor, eq. (\ref{eq_monotonicity}), did not change the results with respect to the FCNN trained without the convexity constraint. This was expected since, as just shown, the data satisfied the convexity checks . We therefore decided to arbitrarily generate non-convex data. Starting with the Ogden fit (Figure \ref{fig_ogden_fit}), additional training data was generated. However, to generate non-convex data, the parameters of the Ogden strain energy were perturbed for different loading curves. Dissipation due to damage was also added for each loading curve. The resulting strain energy contour with respect to the Green Lagrange strain indicates this data is no longer convex (Figure \ref{fig_nonconvex}; Table \ref{table02}). Two FCNNs were trained to these data, one without convexity constraint, and one with the monotonicity test as part of the loss function. The respective strain energy contours are shown in Figure \ref{fig_nonconvex}B and C, and the summary of the tests using eqs. (\ref{eq_convexity_def}) and (\ref{eq_monotonicity}) is given in Table \ref{table02}. Overall, these results indicate that imposing the constraint during training leads to convex strain energies even if the original data is not convex. }

\setlength{\unitlength}{1cm}
\begin{figure}
\centering
\begin{picture}(15,6)
\put(0,0){\includegraphics[width=0.9\linewidth]{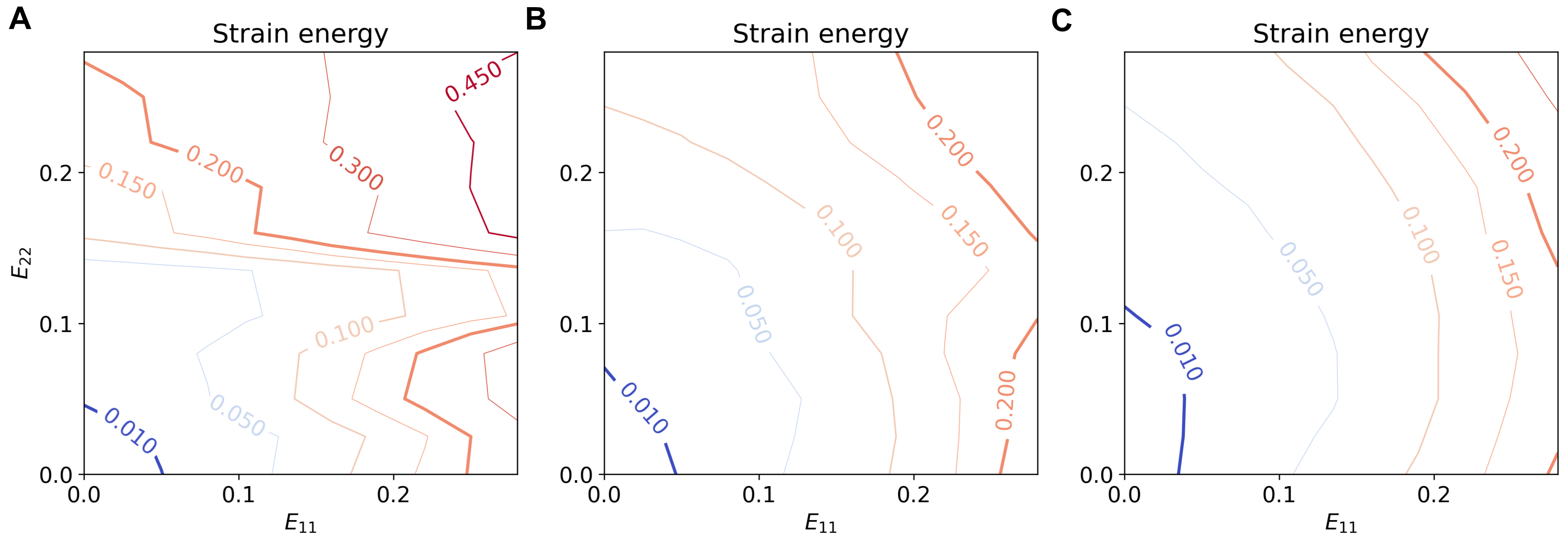} }
\end{picture}
\caption{Non-convex strain energy density function contours (A) generated by perturbing the Ogden fit in Figure \ref{fig_ogden_fit}. A FCNN trained without convexity constraints captures the data and is therefore not convex (B). However, imposing the monotonicity of the second Piola Kirchhoff stress tensor with respect to the right Cauchy Green deformation tensor leads to convex strain energy predictions (C). }
\label{fig_nonconvex} 
\end{figure}

\subsection*{Finite element simulation}

\begin{figure}
\centering
\includegraphics[width=0.95\linewidth]{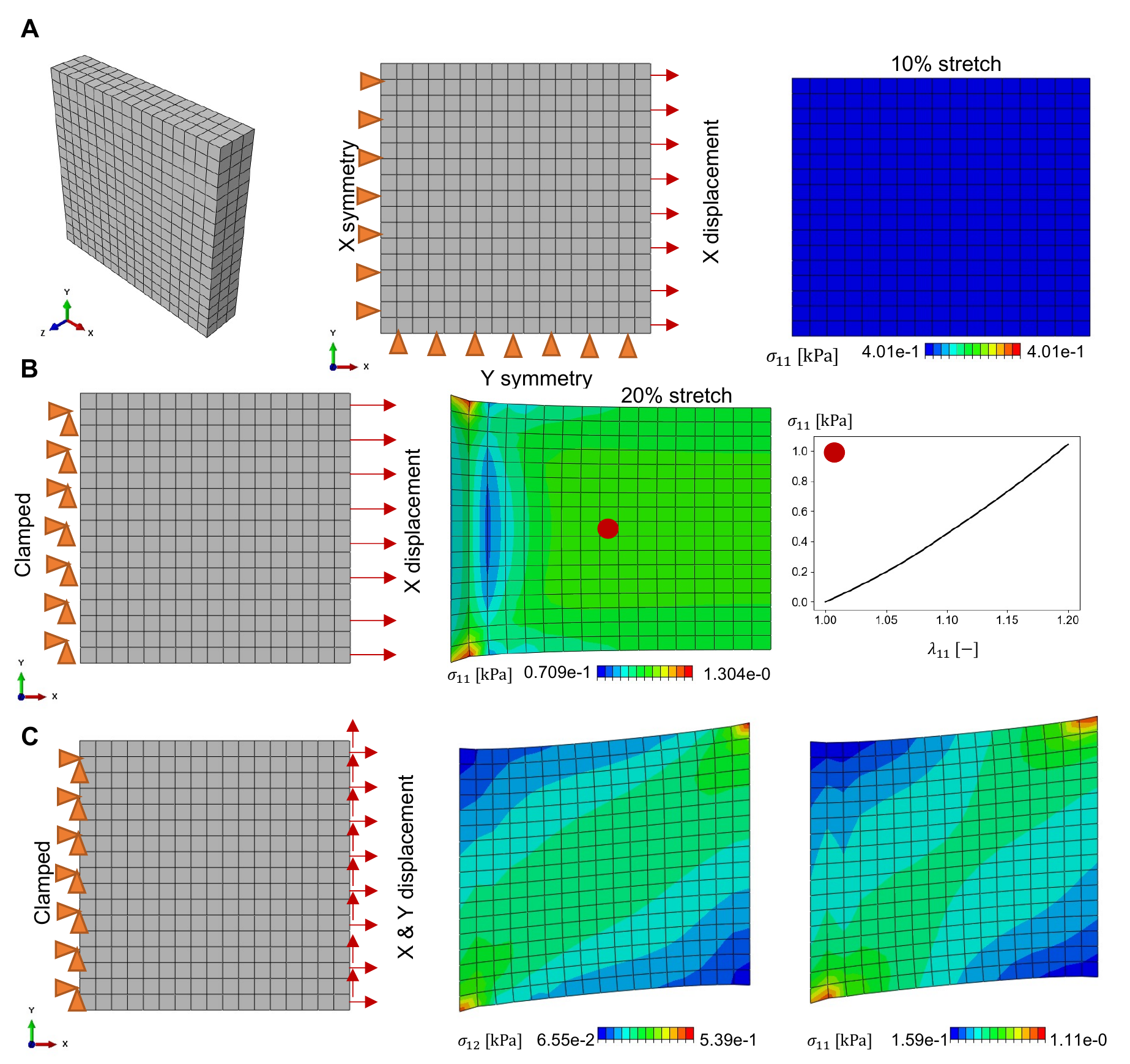}
\caption{ Finite element results using the FCNN in the UMAT subroutine. Uniaxial simulation with symmetry boundary conditions on the left and bottom boundaries leads to a uniform stress field (A). Clamping the left boundary leads to a non-uniform stress distribution; the center of the domain still behaves as in uniaxial loading (B). Shear simulations lead to a band of shear stress (C).}
\label{fig08} 
\end{figure}

Figure \ref{fig08} shows results from three basic tests done to verify the correct implementation of the FCNN in the UMAT subroutine. We first performed  a uniaxial test in which symmetry for X, Y, and Z planes was imposed on the corresponding boundaries. X-displacement was applied on one end of the domain to impose an overall stretch of $\lambda_x=1.1$. The microstructure parameters used corresponded to $\varphi = 0.16 \mu$m and $\theta = 0.45\%$. The result is a homogeneous stress distribution that is required based on the boundary conditions, and it aligns with the FCNN calculation outside of the UMAT, $\sigma_{11}=0.40$ kPa. Secondly, a uniaxial deformation for a higher stretch of $\lambda_x=1.2$ was applied, but in this case, one of the ends was clamped. All other surfaces were traction-free. The finite element simulation converged without problems. In this case, due to the clamped boundary condition on one end, the stress distribution is no longer uniform and boundary effects are visible. The center point of the mesh still behaves correspondingly to uniaxial loading, with a stress that aligns with FCNN predictions. This simulation further suggests that the UMAT was implemented correctly. One last test was performed to verify that the simulation converged under different loading, in particular one that involves shearing. In this third simulation, the left end was kept clamped, but a shear deformation was added on the right boundary. All other surfaces were still traction free. The simulation converged also without a problem, and the resulting stress distribution showed a band of shear and normal stress along the diagonal, in agreement with the overall pattern of stress that would be expected from popular strain energy functions such as the neo-Hookean model. These results suggest that the FCNN was implemented correctly in the UMAT, that the second derivatives obtained through back-propagation are also implemented correctly in the UMAT, and that the potential satisfies convexity leading to the proper convergence of the simulation.

\begin{figure}
\centering
\includegraphics[width=0.95\linewidth]{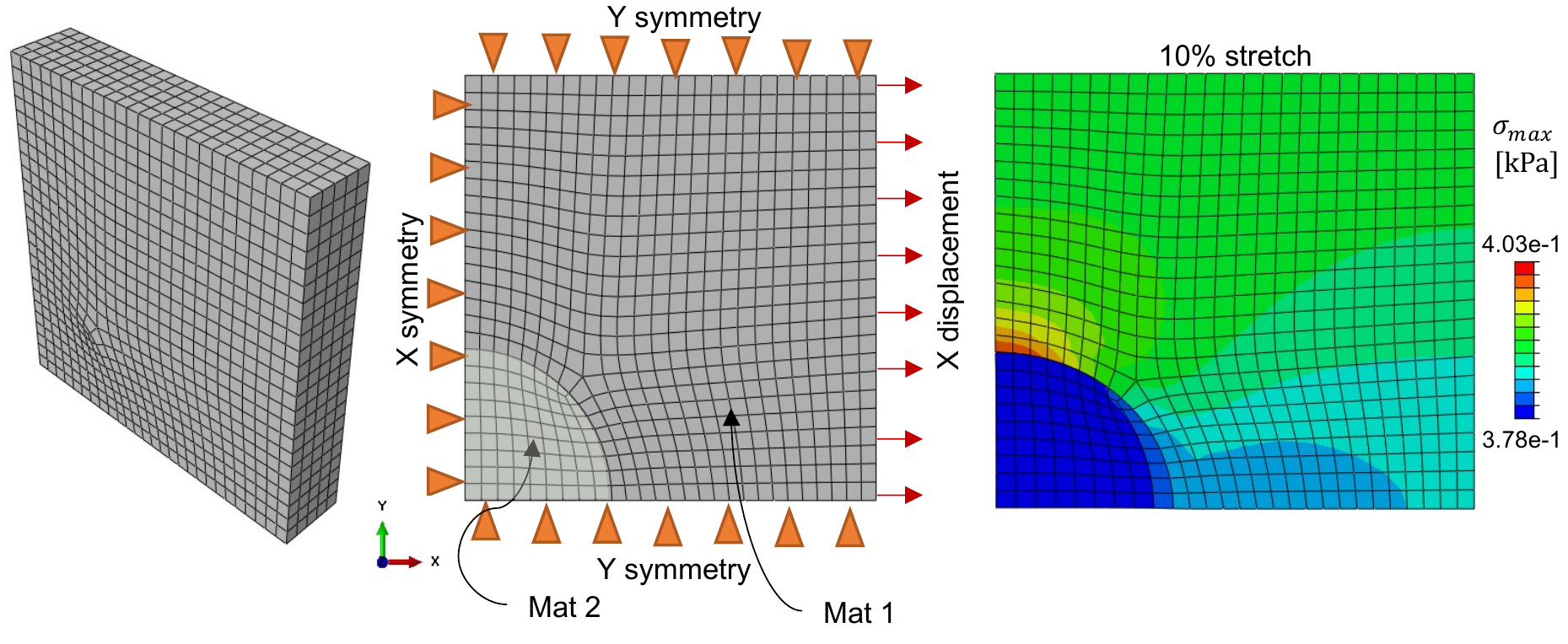}
\caption{ Finite element simulation of a rectangular domain with an inner region made out of a material with microstructure  \textcolor{black}{$\varphi = 0.25 \mu$m and $\theta = 0.75\%$} and a surrounding material defined by \textcolor{black}{$\varphi = 0.1 \mu$m and $\theta = 0.3\%$}. Due to symmetry at the bottom and left, only a quarter of the domain is modeled. The right surface was subjected to uniaxial deformation. Maximum principal stress is shown.}
\label{fig09} 
\end{figure}

One final example was the modeling of a problem more representative of an application of this kind of material model. We modeled a quarter of a domain in which the center was made out of a different material with respect to the surrounding domain. This type of domain could represent a wound or a heterogeneous thrombus depending on the relative material properties. To model only a quarter of the domain, symmetry boundary conditions were used at the bottom and left surfaces, while the top surface has only Y fixed. The right surface was subjected to a uniaxial stretch. The UMAT was used for both domains, but with different inputs for the microstructure variables. \textcolor{black}{For the surrounding material we used $\varphi = 0.1 \mu$m and $\theta = 0.3\%$, while for the interior circular domain the microstructure variables were set to $\varphi = 0.25 \mu$m and $\theta = 0.75\%$.} The simulation results are shown in Figure \ref{fig09}, which shows the maximum principal stress over the entire domain. The microstructure of the inner domain corresponds to softer material behavior of this inclusion, and similar to the Eshelby problem, the stress in the inclusion is nearly constant. The stress in the surrounding stiffer material is higher compared to the inner soft inclusion, \textcolor{black}{and shows the expected stress concentration}. 

\section*{Discussion}

In this study, we proposed to use a FCNN metamodel to replace a microscale fiber network model. The FCNN was trained on network microstructures subjected to a wide set of deformations. While the microscale model takes approximately 1 min to solve for equilibrium of a DFN, the prediction from the FCNN is obtained in less than 1 msec, highlighting the significant speed-up from using machine learning tools to interpolate physics-based models. Significantly, the FCNN model has a simple structure with a small number of parameters and the metamodel can be constrained to satisfy physically meaningful equations, yielding an efficient and numerically stable finite element implementation. \textcolor{black}{The proposed approach is relevant to situations for macroscale finite element simulations of biopolymer gels or soft tissues. In those applications, constitutive models that are invariant, objective, convex, and can output both the stress and the elasticity tensors are needed. Analytical constitutive models at the macroscale are the default option. Here, we created data-driven models that avoid the dependence on a particular functional form of the strain energy. We showed that neural networks trained on post-processed stress-strain data can be used as constitutive models to solve nonlinear equilibrium problems. Additionally, the macroscale response in tissues and biopolymer gels is a function of the microstructure. Thus, it is essential to use information of the microscale mechanics to model the macroscale deformation. A common approach has been to embed or nest models of the microscale into the macroscale simulations. Our work demonstrates that neural networks can be used to homogenize the microscale response, bypassing the need for nested models.}


\textcolor{black}{The FCNN achieved low errors against the validation data, with an average error of 0.27kPa for the $\Psi_1$ output and 0.18 kPa for the $\Psi_2$ output. Relative errors for $\Psi_1$ were 13\% on average, while for $\Psi_2$ they were approximately 170\%. The larger relative error in $\Psi_2$ is due to the low absolute values of $\Psi_2$, which in many cases were close to zero. However, further testing showed that the predictions of the FCNN laid well within the confidence interval for $\Psi_1$ and $\Psi_2$, see Figure \ref{fig05}C,E. Error against the validation data was due in part to the inherent uncertainty in the response of the DFNs.} Even when the homogenized microstructure variables $\theta, \varphi$, were fixed,  the networks were not fully determined. In practice, the networks were generated by seeding a domain with random nucleation points for fiber formation. Hence, this source of randomness led to variance in the mechanical response for multiple networks that had the similar $\theta, \varphi$, \textcolor{black}{as shown in Figure \ref{fig03}}. The type of FCNN used here is a deterministic model. When trained on the data from the different DFNs, the FCNN naturally learns the mean response for a given microstructure. Future work will use Bayesian neural networks that can capture the variance in the response instead of deterministic predictions \cite{mullachery2018bayesian,neal2012bayesian}. Within the validation set, the samples with relatively larger error corresponded to networks with extreme values of microstructure parameters. Moreover, some values of  $\Psi_1$, $\Psi_2$ in the validation set also had outliers for which the FCNN error was larger. These inconsistent values compared to the rest of the training and validation data could be caused by the existence of non-unique solutions of the DFN simulation. Other refinements of the DFN model can be considered. In this effort, we opted for a mechanical equilibrium problem in which fibers contribute only due to stretching but not bending. To make the networks more stable, bending resistance could be considered \cite{licup2015stress,zhang2013microstructural}. Different constitutive models for individual fibers can also be incorporated \cite{aghvami2016fiber,hadi2012simulated,raghupathy2009closed,sander2009image}. 
In addition to the Bayesian approach to handle the variance in the response function, networks with more advanced structure and activation functions can also be considered in the future to improve the training and testing performance while still satisfying the symmetry and convexity requirements \textcolor{black}{\cite{zhang2013microstructural,xu2021learning}}. 

\textcolor{black}{Comparison of the FCNN against analytical strain energies shows that both the FCNN and the Ogden model, but not the neo-Hookean model, can accurately describe the macroscale behavior of biopolymer gels. One advantage of the FCNN approach compared to analytical strain energy functions is that the FCNN uses microstructure information and can predict a wider set of material responses compared to the Ogden model, which has to be fitted to a specific RVE response (see Figure \ref{fig_ogden_fit}). Another advantage of the FCNN approach is that, when considering analytical strain energy functions, the quality of the fit depends on the choice of material model. For example, recent work on mechanical characterization of blood clots investigated three material models, Ogden, Fung and Yeoh, to fit their data \cite{sugerman2021whole}. The FCNN approach bypasses the need to test multiple analytical models and results in accurate material models as demonstrated here.  }

Due to the substantial reduction in computational time and good accuracy in the energy prediction, we embedded our FCNN as a user material \textcolor{black}{subroutine} in the commercial finite element package Abaqus. The finite element implementation showed that the FCNN led to stably convergent simulations. An advantage of our implementation is that the entire FCNN can be passed in a vector of material parameters defined in the input file. In this way, the UMAT does not evaluate the specific FCNN trained with the microscale mechanics data and can actually be used to evaluate any FCNN specified in the input file. For example the recent work by \cite{liu2020generic} shows that a FCNN can be used to interpolate heart valve biaxial data. The metamodel in \cite{liu2020generic} was not used in finite element simulations, however, due to the generality of our UMAT, the FCNN in \cite{liu2020generic} could be easily evaluated within our UMAT.  All the files for this manuscript and a video tutorial to use our UMAT are available online.

The work presented here is informed by previous work on fibrin and collagen network modeling as well as experimental characterization of biopolymer gel mechanics \cite{aghvami2016fiber,de2016combined,rudnicki2013nonlinear,wang2021probing, kumar2020structural}. Nevertheless, future experimental work by our group will further explore the multiscale mechanics of fibrin and collagen gels to fully calibrate and validate our computational model. \textcolor{black}{Additionally, the proposed methodology is currently restricted to fully incompressible isotropic hyperelasticity, and the data for training is restricted to plane stress tests. Future work will extend the framework to include anisotropy, compressibility, and dissipative effects (e.g. viscoelastic response), as well as extend the training data to include arbitrary deformations and not just plane stress. }

\section*{Conclusions}
Our work demonstrates that neural networks can be trained by micromechanical simulations, which capture ECM network behaviors and their relation to microstructural variables such as volume fraction and fiber diameter. The computational efficiency of this kind of machine learning metamodel makes it suitable for implementation within finite element simulations. More importantly, the FCNN relies on the data of the microscale model and not on any sort of analytical approximation. Future work includes using the FCNN to predict rate-dependent constitutive relations and for inverse estimation of microstructure parameters from a combination of mechanical tests and imaging data.

\section*{Acknowledgements}

This work was supported in part by the National Science
Foundation under grant No. 1911346-CMMI to PIs Tepole and Calve and the Bilsland Dissertation Fellowship to Dr. Yue Leng. 
All the files associated with this publication are available at \url{https://bitbucket.org/buganzalab/nn_rve}, and a video tutorial to use our UMAT is \url{available at https://www.youtube.com/watch?v=kdXgnZfSMSg&t=23s}.

%






\end{document}